\documentclass{llncs}

\title{``Almost-stable'' matchings in the \\ Hospitals / Residents problem with Couples}

\author{David F. Manlove\thanks{Supported by Engineering and Physical Sciences Research Council grants EP/K010042/1 and EP/N508792/1.  Email {\tt david.manlove@glasgow.ac.uk}.}, Iain McBride\thanks{Supported by a SICSA Prize PhD Studentship.} and James Trimble$^{*}$}

\institute{School of Computing Science, Sir Alwyn Williams Building, \\ University of Glasgow, Glasgow G12 8QQ, UK} 

\usepackage{makeidx}
\usepackage{hyperref}
\usepackage{pdflscape}
\usepackage{amsmath}
\usepackage{amsfonts}
\usepackage{verbatim}
\usepackage{graphicx}
\usepackage{algorithm}
\usepackage{algorithmic}
\usepackage{textcomp}
\usepackage{float}
\usepackage{epstopdf}
\usepackage{comment}

\usepackage{pdfpages}
\usepackage{tabulary}
\usepackage{cite}

\newtheorem{defn}{Definition}
\newtheorem{lemma1}[defn]{Lemma}
\newtheorem{theorem1}[defn]{Theorem}
\newtheorem{proposition1}[defn]{Proposition}
\newtheorem{corollary1}[defn]{Corollary}
\begin{document}
\frontmatter          % for the preliminaries
\maketitle
\pagestyle{headings}
\pagenumbering{arabic}
%----------------------------------------------------------------------------------------------------------------------------------------------------
\begin{abstract}
%The Hospitals / Residents problem with Couples ({\sc hrc}) is a generalisation of the classical Hospitals / Residents problem ({\sc hr}) that is important in practical applications because it models the case where couples submit joint preference lists over pairs of (typically geographically close) hospitals. In this paper we give a new NP-completeness result for the problem of deciding whether a stable matching exists, in highly restricted instances of {\sc hrc}, and also show that we can decide in polynomial time whether a stable matching exists in an instance of {\sc hrc} in which each couple's joint preference list is of length exactly one and the single residents' and hospitals' preference lists are of length two. Further, we present a full description of the first Integer Programming model for finding a maximum cardinality most-stable matching in an instance of {\sc hrc} and we describe empirical results when this model applied to randomly generated instances of {\sc hrc} reflecting the properties of the instances arising in real world allocation problems in Scotland.
The Hospitals / Residents problem with Couples ({\sc hrc}) models the allocation of intending junior doctors to hospitals where couples are allowed to submit joint preference lists over pairs of (typically geographically close) hospitals.  It is known that a stable matching need not exist, so we consider {\sc min bp hrc}, the problem of finding a matching that admits the minimum number of blocking pairs (i.e., is ``as stable as possible'').  We show that this problem is NP-hard and difficult to approximate even in the highly restricted case that each couple finds only one hospital pair acceptable.  However if we further assume that the preference list of each single resident and hospital is of length at most 2, we give a polynomial-time algorithm for this case.  We then present the first Integer Programming (IP) and Constraint Programming (CP) models for {\sc min bp hrc}.  Finally, we discuss an empirical evaluation of these models applied to randomly-generated instances of {\sc min bp hrc}.  We find that on average, the CP model is about 1.15 times faster than the IP model, and when presolving is applied to the CP model, it is on average 8.14 times faster.  We further observe that the number of blocking pairs admitted by a solution is very small, i.e., usually at most 1, and never more than 2, for the (28,000) instances considered.
\end{abstract}

%----------------------------------------------------------------------------------------------------------------------------------------------------

\section{Introduction}
\vspace{-1mm}
\label{section:introduction}
\noindent \textbf{The Hospitals / Residents problem.} The \emph{Hospitals / Residents problem} ({\sc hr}) \cite{GS62} is a many-to-one allocation problem that models the assignment process involved in centralised matching schemes such as the National Resident Matching Program (NRMP) \cite{ZZZ5} which assigns graduating medical students to hospital posts in the USA. Analogous schemes exist in Canada \cite{ZZZ6} and Japan \cite{ZZZ50}. A similar process was used until recently to match medical graduates to Foundation Programme places in Scotland: the Scottish Foundation Allocation Scheme (SFAS) \cite{Irv98}. Moreover, similar matching schemes exist in the context of Higher Education admission in Hungary \cite{Bir08, ZZZ53}, Spain \cite{Rom98}, Turkey \cite{BS99} and Ireland \cite{ZZZ9, ZZZ53}. The reader is referred to \cite{ZZZ53} for details of matching practices in a number of practical contexts throughout Europe.

An instance of {\sc hr} consists of two sets of agents -- a set $R= \{r_1 ,\ldots r_{n_1}\}$ containing \emph{residents} and a set $H= \{h_1 ,\ldots h_{n_2}\}$ containing \emph{hospitals}. Every resident expresses a linear preference over some subset of the hospitals, his \emph{preference list}. The hospitals in a resident's preference list are his \emph{acceptable} partners; all other hospitals being \emph{unacceptable}. Every hospital expresses a linear preference over those residents who find it acceptable. Further, each hospital $h_j\in H$ has a positive integral \emph{capacity} $c_j$, the maximum number of residents to which it may be assigned. A \emph{matching} $M$ is a set of acceptable resident-hospital pairs such that each resident appears in at most one pair and each hospital $h_j$ belongs to at most $c_j$ pairs. If $(r_i,h_j)\in M$ then $r_i$ is said to be \emph{assigned} to $h_j$, $M(r_i)$ denotes $h_j$, and $r_i$ is an \emph{assignee} of $h_j$.  Given $r_i\in R$, if $r_i$ does not belong to any pair in $M$ then $r_i$ is said to be \emph{unassigned}.  % and we define $M(r_i)=\emptyset$.  
Given $h_j\in H$, we let $M(h_j)$ denote the set of assignees of $h_j$ in $M$.  Hospital $h_j$ is \emph{undersubscribed}, \emph{full} or \emph{oversubscribed} according as $|M(h_j)$ is less than, equal to, or larger than $c_j$, respectively.

Roth \cite{Rot84} argued that a key property to be satisfied by any matching $M$ in an instance $I$ of {\sc hr} is \emph{stability}, which ensures that $M$ admits no \emph{blocking pair} in $I$.  Informally, such a pair comprises a resident $r_i$ and a hospital $h_j$, both of whom have an incentive to disregard their assignments (if any) and become matched to one another outside of $M$, undermining its integrity.
A matching is \emph{stable} if it admits no blocking pair. It is known that every instance of {\sc hr} admits at least one stable matching, which can be found in time linear in the size of the instance \cite{GS62}. %Moreover, for an arbitrary {\sc hr} instance $I$, any resident assigned in one stable matching in $I$ is assigned in all stable matchings in $I$. Further, any hospital undersubscribed in some stable matching in $I$ is assigned exactly the same set of residents in every stable matching in $I$ \cite{GS85, Rot86, Rot84}. 

%The size of these matching programmes can vary tremendously: the SFAS programme typically involved approximately 750 applicants and 50 hospitals annually; in the Hungarian Higher Education allocation process in 2011, 140,953 applicants participated \cite{ZZZ51}; in 2015 the NRMP process involved 52,880 residents and assigned 26,252 residents over 27,293 posts \cite{NRMP2015}; and in the case of the largest centralised allocation scheme currently known, the Chinese Higher Education matching scheme involved approximately ten million applicants in 2007 \cite{Zha09}. Clearly the efficiency with which an optimal matching might be obtained is of great practical interest. 

%\subsubsection{The Hospitals / Residents problem with Couples}
%\smallskip
\medskip
\noindent \textbf{The Hospitals / Residents problem with Couples.} The Hospitals / Residents problem with Couples ({\sc hrc}) is a generalisation of {\sc hr} that is important in practical applications because it models the case where some of the residents may apply jointly in couples, so that they may be matched to hospitals that are geographically close to one another.  In order to ensure this, a couple submits a joint preference list over pairs of hospitals, rather than individual hospitals.  Matching schemes for junior doctors such as the NRMP \cite{ZZZ5} allow couples to apply jointly, as do assignment processes in the US Navy \cite{Sho00,Rob01,YGS03} (for which {\sc hrc} is an appropriate problem model), for example.

Formally, an instance $I$ of {\sc hrc} consists of a set $R=\{r_1 ,\ldots r_{n_1}\}$ containing \emph{residents} and a set $H=\{h_1 ,\ldots h_{n_2}\}$ containing \emph{hospitals}. The residents in $R$ are partitioned into two sets, $S$ and $S^{ \prime}$. The set $S$ consists of \emph{single} residents and the set $S^{\prime }$ consists of those residents involved in \emph{couples}. There is a set $C = \{(r_i, r_j): r_i, r_j \in S^{\prime}\}$ of \emph{couples} such that each resident in $S^{\prime}$ belongs to exactly one pair in $C$.

Each single resident $r_i \in S$ expresses a linear preference order over some subset of the hospitals, his \emph{acceptable} hospitals; all other hospitals being \emph{unacceptable}. Each couple $(r_i, r_j)\in C$ expresses a joint linear preference order over a subset $A$ of $H \times H$ where $(h_p, h_q)\in A$ represents the simultaneous assignment of $r_i$ to $h_p$ and $r_j$ to $h_q$. The hospital pairs in $A$ represent those joint assignments that are \emph{acceptable} to $(r_i, r_j)$, all other joint assignments being \emph{unacceptable}. Each hospital $h_j \in H$ expresses a linear preference order over those residents who find it acceptable, either as a single resident or as part of a couple, and as in the case of {\sc hr}, each hospital $h_j \in H$ has a positive integral \emph{capacity} $c_j$. 

A \emph{matching} $M$ in $I$ is defined as in {\sc hr} case, with the additional restriction that, for each couple $(r_i, r_j)\in C$, either both $r_i$ and $r_j$ appear in no pair of $M$, or else $\{(r_i,h_k),(r_j,h_l)\}\subseteq M$ for some pair $(h_k,h_l)$ that $(r_i,r_j)$ find acceptable.  In the former case, $(r_i,r_j)$ are said to be \emph{unassigned}, whilst in the latter case, $(r_i,r_j)$ are said to be \emph{jointly assigned} to $(h_k,h_l)$.  Given a resident $r_i\in R$, the definitions of $M(r_i)$, \emph{assigned} and \emph{unassigned} are the same as for the {\sc hr} case, whilst for a hospital $h_j\in H$, the definitions of \emph{assignees}, $M(h_j)$, \emph{undersubscribed}, \emph{full} and \emph{oversubscribed} for hospitals are also the same as before.

We seek a \emph{stable} matching, which guarantees that no resident and hospital, and no couple and pair of hospitals, have an incentive to deviate from their assignments and become assigned to each other outside of the matching.  Roth \cite{Rot84} considered stability in the {\sc hrc} context but did not define the concept explicitly. Whilst Gusfield and Irving \cite{GI89} gave a formal definition of a blocking pair, it neglected to deal with the case that both members of a couple may wish to be assigned to the same hospital.  A number of other stability definitions for {\sc hrc} have since been given in the literature that address this issue (see \cite{BK13} and \cite[Section 5.3]{Man13} for more details), including that of McDermid and Manlove \cite{MM10}, which we adopt in this paper.  We repeat their definition again here for completeness.  %In what follows 
%extended their definition to deal with this possibility (however both definitions are equivalent in the case that no pair of the form $(h_p , h_p)$ appears in any couple's preference list).   We adopt Manlove and McDermid's stability definition in this paper, and repeat the definition again here.
%Given an instance $I$ of {\sc hrc} and a matching $M$ in $I$, we denote $r_i$'s assigned partner in $M$ by $M(r_i)$. Similarly, we denote the set of $h_j$'s assigned partners in $M$ by $M(r_j)$.

\begin{defn}[\hspace{-0.1mm}\cite{MM10}]
\label{stability:MM}
Let $I$ be an instance of {\sc hrc}.  A matching $M$ is \emph{stable} in $I$ if none of the following holds:  
\begin{enumerate}
\item There is a single resident $r_i$ and a hospital $h_j$, where $r_i$ finds $h_j$ acceptable, such that either $r_i$ is unassigned in $M$ or prefers $h_j$ to $M(r_i$), and either $h_j$ is undersubscribed in $M$ or prefers $r_i$ to some member of $M(h_j)$.
\item There is couple $(r_i, r_j)$ and a hospital $h_k$ such that \emph{either}
\begin{enumerate}
\item[(a)] $(r_i, r_j)$ prefers $(h_k, M(r_j))$ to $(M(r_i), M(r_j))$, and either $h_k$ is  undersubscribed in $M$ or prefers $r_i$ to some member of $M(h_k)\backslash \{r_j\}$ \emph{or}
\item[(b)] $(r_i, r_j)$ prefers $(M(r_i), h_k)$ to $(M(r_i), M(r_j))$, and either $h_k$ is undersubscribed in $M$ or prefers $r_j$ to some member of $M(h_k)\backslash \{r_i\}$.
\end{enumerate}
\item There is a couple $(r_i, r_j)$ and a pair of (not necessarily distinct) hospitals $h_k\neq M(r_i)$, $h_l\neq M(r_j)$ such that $(r_i,r_j)$ finds $(h_k,h_l)$ acceptable, and either $(r_i,r_j)$ is unassigned or prefers the joint assignment $(h_k, h_l)$ to $(M(r_i), M(r_j))$, and \emph{either}
\begin{enumerate}
\item[(a)] $h_k\neq h_l$, and $h_k$ (respectively $h_l$) is either undersubscribed in $M$ or prefers $r_i$ (respectively $r_j$) to at least one of its assignees in $M$; \emph{or}
\item[(b)] $h_k=h_l$, and $h_k$ has two free posts in $M$, i.e., $c_k-|M(h_k)|\geq 2$; \emph{or}
\item[(c)] $h_k=h_l$, and $h_k$ has one free post in $M$, i.e., $c_k-|M(h_k)|=1$, and $h_k$ prefers at least one of $r_i,r_j$ to some member of $M(h_k)$; \emph{or}
\item[(d)] $h_k=h_l$, $h_k$ is full in $M$, $h_k$ prefers $r_i$ to some $r_s\in M(h_k)$, and $h_k$ prefers $r_j$ to some $r_t\in M(h_k)\backslash \{r_s\}$.
\end{enumerate}

\medskip
A resident and hospital, or a couple and hospital pair, satisfying one of the above conditions, is called a \emph{blocking pair} of $M$ and is said to \emph{block} $M$.
%We may also use the terminology \emph{Type $X$-blocking pair} depending on which condition is satisfied.
\end{enumerate}
%\caption{The Manlove-McDermid Stability definition. \cite{MM10}}
\end{defn}

\noindent \textbf{Existing algorithmic results for {\sc hrc}.} An instance $I$ of {\sc hrc} need not admit a stable matching \cite{Rot84}.  We call $I$ \emph{solvable} if it admits a stable matching, and \emph{unsolvable} otherwise.  Also an instance of {\sc hrc} may admit stable matchings of differing sizes \cite{AC96}.  Further, the problem of deciding whether a stable matching exists in an instance of {\sc hrc} is NP-complete, even in the restricted case where there are no single residents and each hospital has capacity 1 \cite{NH88,Ron90}.  The decision problem is also W[1]-hard \cite{MS11} when parameterized by the number of couples.

In many practical applications of {\sc hrc} the residents' preference lists are short. Let $(\alpha, \beta, \gamma)$-{\sc hrc} denote the restriction of {\sc hrc} in which each single resident's preference list contains at most $\alpha $ hospitals, each couple's preference list contains at most $\beta $ pairs of hospitals and each hospital's preference list contains at most $\gamma $ residents.  Bir\'o et al.\ \cite{BMMcB14} showed that deciding whether an instance of \small $(0,2,2)$\normalsize {\sc -hrc} admits a stable matching is NP-complete.

Heuristics for {\sc hrc} were described and compared experimentally by Bir\'o et al.\ \cite{BIS11}.  As far as exact algorithms are concerned, Bir\'o et al.\ \cite{BMMcB14} gave an Integer Programming (IP) formulation for finding a maximum cardinality stable matching (or reporting that none exists) in an arbitrary instance of {\sc hrc} and presented an empirical evaluation of an implementation of their model, showing that their formulation was capable of solving instances of the magnitude of those arising in the SFAS application.  Further algorithmic results for {\sc hrc} are given in \cite{BK13,Man13,McB15}. 

\medskip
\noindent \textbf{Most-stable matchings.} Given that a stable matching need not exist in a given {\sc hrc} instance $I$, a natural question to ask is whether there is some other matching that might be the best alternative amongst the matchings in $I$. Roth \cite{Rot90, Rot91} argued that instability in the outcome of an allocation process gives participants a greater incentive to circumvent formal procedures; it follows minimising the amount of instability might be a desirable objective.  Eriksson and H\"aggstr\"om \cite{EH08} suggested that the number of blocking pairs admitted by a matching is a meaningful way to measure its degree of instability.

Define $bp(M)$ to be the set of blocking pairs relative to a matching $M$ in $I$, and define a \emph{most-stable matching} to be a matching $M$ for which $|bp(M)|$ is minimum, taken over all matchings in $I$.  Clearly if $I$ admits a stable matching $M$, then $M$ is a most-stable matching in $I$.  Let {\sc min bp hrc} denote the problem of finding a most-stable matching, given an instance of {\sc hrc}.  Most-stable matchings have been studied from an algorithmic point of view in various matching problem contexts \cite{ABM06,BMM10,BMM12,FKPS10,HIM09,HIM16} (see \cite{Man13} for more details), including in humanitarian organisations \cite{Sol14}.  Define \small $(\alpha ,\beta, \gamma)$\normalsize {\sc -min bp hrc} to be the restriction of {\sc min bp hrc} to instances of \small $(\alpha ,\beta, \gamma)$\normalsize {\sc -hrc}. 

\medskip
%\smallskip
\noindent \textbf{Contribution of this work.}  In Section \ref{section:complexitySEA} we show that $(\infty,1,\infty)$\normalsize {\sc -min bp hrc} is NP-hard and not approximable within $n_1^{1-\varepsilon}$, for any $\varepsilon>0$, unless P=NP (recall that $n_1$ is the number of residents in a given instance).  In this highly restricted case of {\sc min bp hrc}, each couple finds only one hospital pair acceptable and each hospital has capacity 1 ($\infty$ refers to preference lists of unbounded length).  We also show that $(\infty,\infty,1)$\normalsize {\sc -min bp hrc} and $(2,1,2)$\normalsize {\sc -min bp hrc} are solvable in polynomial time.  These results help to narrow down the search for the boundary between polynomial-time solvable and NP-hard restrictions of {\sc min bp hrc} (recall that $(0,2,2)$\normalsize {\sc -min bp hrc} is NP-hard \cite{BMMcB14}).

In Section \ref{section:IPModels_MINBPHRC_SEA} we present the first IP model for {\sc min bp hrc}; indeed this model can be used to find a most-stable matching of maximum cardinality.  This formulation extends our earlier IP model for {\sc hrc}, presented in \cite{BMMcB14}.  Then in Section \ref{section:MIN_BP_HRC_IPexperiments} we present data from an empirical evaluation of an implementation of the IP model for {\sc min bp hrc} applied to randomly-generated instances.  We measure the mean solution time, mean size of a most-stable matching and mean number of blocking pairs admitted by a most-stable matching when varying (i) the number of residents, (ii) the number of couples, (iii) the number of hospitals and (iv) the lengths of the residents' preference lists.  Our main finding is that, over the 28,000 instances considered, the number of blocking pairs admitted by a most-stable matching is very small: it is usually at most 1, and never more than 2.  This suggests that in a given {\sc hrc} instance in practice, even if a stable matching does not exist, we may be able to find a matching with only a very small amount of instability.

Finally, in Section \ref{section:CPmodel} we present the first Constraint Programming (CP) model for {\sc min bp hrc} and evaluate its performance compared to the IP model over the instances used for the empirical analysis in Section \ref{section:MIN_BP_HRC_IPexperiments}.  We observe that on average, the CP model is about 1.15 times faster than the IP model, and when presolving is applied to the CP model, it is on average 8.14 times faster. 

\medskip
\noindent {\bf Related work.} Drummond et al.\ \cite{DPB15} presented SAT and IP encodings of {\sc hrc} and investigated empirically their performance, along with two earlier heuristics for {\sc hrc}, on randomly-generated instances.  Their main aim was to measure the time taken to find a stable matching or report that none exists, and the proportion of solvable instances.  They found that the SAT encoding gave the fastest method and was generally able to resolve the solvability question for the highest proportion of instances.  In another paper \cite{PDB15}, the same authors conducted further empirical investigations on random instances using an extension of their SAT encoding to determine how many stable matchings were admitted, and whether a resident Pareto optimal stable matching existed.  We remark that the results in \cite{DPB15,PDB15} are not directly comparable to ours, because the stability definition considered in those papers is slightly weaker than that given by Definition \ref{stability:MM}.  See Appendix \ref{sec:stabdefcomp} for a discussion of this issue.
%Bstriction of {\sc hrc} in which the capacities of the hospitals may not be greater than one. They found that the SAT encoding was more successful than the IP encoding at determining if an instance of {\sc hrc} did not admit a stable matching. 
%Drummond et al.\ \cite{DPB15} described a notion of stability that is very closely related Definition \ref{stability:MM}. In fact, for a given instance $I$ of {\sc hrc}, the set of stable matchings admitted by $I$ under Drummond et al.\'s definition of stability is a subset of the set of matchings that are stable according to Definition \ref{stability:MM}.

Hinder \cite{Hin15} presented an IP model for a general stable matching problem with contracts, which includes {\sc hrc} as defined here, as a special case.  He conducted an empircal study on randomly-generated instances, comparing the performance of the IP model, its LP relaxation and a previously-published heuristic.  Hinder showed that the LP relaxation finds stable matchings (when they exist) with much higher probability than the heuristic, and with probability quite close to the true value given by the IP model.  The IP model terminates surprisingly quickly when the number of residents belonging to a couple is 10\%, but it should be emphasised that in Hinder's random instances, all hospitals have capacity 1.  In such a case our IP/CP models would be much simpler and need not involve the constraints corresponding to stability criteria 3(b), 3(c) and 3(d) in Definition \ref{stability:MM}, thus our runtime results are not directly comparable to Hinder's.

To the best of our knowledge there have been no previous CP models for {\sc hrc}, though a CP model for {\sc hr} was given in \cite{MOPU07}, extending an earlier CP model for the classical Stable Marriage problem, the 1-1 restriction of {\sc hr} \cite{GIMPS01}.  A detailed survey of CP models for stable matching problems is given in \cite[Section 2.5]{Man13}.

Nguyen and Vohra \cite{NV14} proved a remarkable result, namely that it is always possible to find a stable matching in an instance of {\sc hrc} if the capacity of each hospital can be adjusted (up or down) by at most 4, with the total capacity of the hospitals increasing by at most 9. 

\section{Complexity results for {\sc min bp hrc}}
\label{section:complexitySEA}
In this section we present complexity and approximability results for {\sc min bp hrc} in the case that preference lists of some or all of the agents are of bounded length.  We begin with \small$(\infty ,1, \infty)$\normalsize {\sc -min bp hrc}, the restriction in which each couple lists only one hospital pair on their preference list.  Even in this highly restricted case, the problem of finding a most-stable matching is NP-hard and difficult to approximate.  The proof of this result, given in Appendix \ref{sec:inapprox}, begins by showing that, given an instance of \small$(\infty ,1, \infty)$\normalsize {\sc -hrc},  the problem of deciding whether a stable matching exists is NP-complete.  Then a gap-introducing reduction is given from this problem to \small$(\infty ,1, \infty)$\normalsize {\sc -min bp hrc}.
\begin{theorem1}
%\label{inapprox}
\small$(\infty ,1, \infty)$\normalsize {\sc -min bp hrc} is NP-hard and not approximable within a factor of $n_1^{1-\varepsilon}$, for any $\varepsilon>0$, unless P=NP, where $n_1$ is the number of residents in a given instance.  The result holds even if each hospital has capacity 1.
\end{theorem1}

We now turn to the case that hospitals' lists are of bounded length.  It will be helpful to introduce the notion of a \emph{fixed assignment} in a given {\sc hrc} instance $I$.  This involves either (i) a resident-hospital pair $(r_i,h_j)$ such that $h_j$ is the first choice of $r_i$, and $r_i$ is among the first $c_j$ choices of $h_j$, or (ii) a pair comprising a couple $(r_i,r_j)$ and a pair of hospitals $(h_p,h_q)$ such that $h_p$ (resp.\ $h_q$) is the first choice of $r_i$ (resp.\ $r_j$), and $r_i$ (resp.\ $r_j$) is among the first $c_p$ (resp.\ $c_q$) choices of $h_p$ (resp.\ $h_q$).  Clearly any stable matching must contain all the fixed assignments in $I$.  By eliminating the fixed assignments iteratively, we arrive at the following straightforward result for \small$(\infty,\infty,1)$\normalsize {\sc -hrc} (the proofs of all the results stated in this section from this point onwards can be found in Appendix \ref{sec:algs}).
\begin{proposition1}
%\label{proposition:221hrc}
An instance $I$ of \small$(\infty , \infty , 1)$\normalsize {\sc -hrc} admits exactly one stable matching, which can be found in polynomial time.
\end{proposition1}

We now consider the \small$(2,1,2)$\normalsize {\sc -hrc} case.  The process of \emph{satisfying} a fixed assignment involves matching together the resident(s) and hospital(s) involved, deleting the agents themselves (and removing them from the remaining preference lists).  This may uncover further fixed assignments, which themselves can be satisfied.  Once this process terminates, we say that all fixed assignments have been \emph{iteratively satisfied}.  Let $I$ be the \small $(2,1,2)$\normalsize {\sc -hrc} instance that remains.  It turns out that $I$ has a special structure, as the following result indicates.
\begin{lemma1}
\label{lem1}
An arbitrary instance of {\small$(2,1,2)$\normalsize {\sc -hrc}} involving at least one couple and in which all fixed assignments have been iteratively satisfied must be constructed from sub-instances of the form shown in Figure \ref{preflists:the full instance} (see Appendix \ref{sec:algs}) in which all of the hospitals have capacity 1.
\end{lemma1}
It is then straightforward to find a most-stable matching in each such sub-instance.
\begin{lemma1}
\label{lem2}
Let $I'$ be an instance of \small$(2,1,2)$\normalsize {\sc -hrc} of the form shown in Figure \ref{preflists:the full instance} in Appendix \ref{sec:algs}.  If $I'$ has an even number of couples then $I'$ admits a stable matching $M$.  Otherwise $I'$ admits a matching $M$ such that $|bp(M)|=1$ in $I'$.
\end{lemma1}
Using Lemmas \ref{lem1} and \ref{lem2}, it follows that we can find a most-stable matching in an instance $I$ of {\small$(2,1,2)$\normalsize {\sc -hrc}} as follows.  Assume that $M_0$ is the matching in $I$ in which all fixed assignments have been iteratively satisfied, and assume that the corresponding deletions have been made from the preference lists in $I$, yielding instance $I'$.  Lemma \ref{lem1} shows that $I'$ is a union of disjoint sub-instances $I_1,I_2,\dots,I_t$, where each $I_j$ is of the form shown in Figure \ref{preflists:the full instance} in Appendix \ref{sec:algs} ($1\leq j\leq t$).  Let $j$ ($1\leq j\leq t$) be given and let $N_j$ be the number of couples in $I_j$.  Lemma \ref{lem2} implies that, if $N_j$ is even, we may find a stable matching $M_j$ in $I_j$, otherwise we may find a matching $M_j$ in $I_j$ such that $|bp(M_j)|=1$ in $I_j$.  It follows that $M=\cup_{j=0}^t M_j$ is a most-stable matching in $I$.  This leads to the following result.
\begin{theorem1}
\small$(2,1,2)$\normalsize {\sc -min bp hrc} is solvable in polynomial time.
%we can find a maximum cardinality stable matching or report that none exists in polynomial time.
%\label{212HRCEfficientAlgorithmExists}
\end{theorem1}

It remains open to resolve the complexity of \small$(p,1,q)$\normalsize {\sc -hrc} for constant values of $p$ and $q$ where $\max\{p,q\}\geq 3$.

%We have shown that the problem of deciding whether a stable matching exists is NP-complete even for instances of \small$(\infty ,1, \infty)$\normalsize {\sc -hrc} and \small$(2,2)$\normalsize {\sc -hrc} and we presented a polynomial-time algorithm for \small$(2 ,1, 2)$\normalsize {\sc -hrc}. However, the complexity of \small$(2 ,1, 3)$\normalsize {\sc -hrc} and  \small$(3 ,1, 2)$\normalsize {\sc -hrc} remains open.

\section{An Integer Programming formulation for {\sc min bp hrc}}

\label{section:IPModels_MINBPHRC_SEA}
In this section we describe our IP model for {\sc min bp hrc}, which extends the earlier IP model for {\sc hrc} presented in \cite{BMMcB14} (we discuss relationships between the two models at the end of this section).  Let $I$ be an instance of {\sc hrc}; we will denote by $J$ the IP model corresponding to $I$.  Due to space limitations we will only present some of the constraints in $J$; the full description of $J$ is contained in Appendix \ref{section:IPModels_MINBPHRC}.
%and let $J$ be corresponding IP model as described in that paper.  We will show how to modify $J$ in order to model the {\sc min bp hrc} problem.

\medskip
\noindent {\bf Notation.}  We first define some required notation in $I$.
%Let $R = \{r_1, r_2,\dots, r_{n_1}\}$ be the set of residents and let $H=\{h_1, h_2,\dots, h_{n_2}\}$ be the set of hospitals in $I$.
Without loss of generality, suppose residents $r_1, r_2\ldots r_{2c}$ are in couples.  Thus $r_{2c+1}, r_{2c+2}\ldots r_{n_1}$ comprise the single residents.  %Each resident $r_i\in R$ has a preference list of length $l(r_i)$ consisting of individual hospitals $h_j\in H$. %Each hospital $h_j\in H$ has a preference list of individual residents $r_i\in R$ of length $l(h_j)$. Further, 
%(The individual preference lists of residents involved in a couple are combined in a precise way -- see \cite{BMMcB14} for further details.)
%Also each hospital $h_j\in H$ has capacity $c_j \geq 1$, the number of residents with which it may match.
%Without loss of generality, suppose residents $r_1, r_2\ldots r_{2c}$ are in couples.
Again, without loss of generality, suppose that the couples are $(r_{2i-1}, r_{2i})$  $(1\leq i\leq c)$.
Suppose that the joint preference list of  a couple $\mathcal C_i =  (r_{2i-1}, r_{2i})$ is
$(h_{\alpha_1}, h_{\beta_1}),(h_{\alpha_2}, h_{\beta_2})\ldots (h_{\alpha_l}, h_{\beta_l})$.  From this list we say that
$h_{\alpha_1}, h_{\alpha_2}\ldots h_{\alpha_l}$ and $h_{\beta_1}, h_{\beta_2}\ldots h_{\beta_l}$ are the \emph{individual} preference lists for $r_{2i-1}$ and  $r_{2i}$ respectively.
%Clearly, the projected preference list of the residents $r_{2i-1}$ and $r_{2i}$ are the same length as the preference list of the couple $\mathcal C_i = (r_{2i-1}, r_{2i})$.
%Let $l(\mathcal C_i)$ denote the length of the preference list of $\mathcal C_i$ and let $l(r_{2i-1})$ and $l(r_{2i})$ denote the lengths of the individual preference lists of $r_{2i-1}$ and $r_{2i}$ respectively. Clearly we have that $l(r_{2i-1}) = l(r_{2i}) = l(\mathcal C_i)$. A given hospital $h_j$ may appear more than once in the individual preference list of a resident belonging to a couple $\mathcal C_i = (r_{2i-1}, r_{2i})$.  Similarly
Let $l(r_i)$ denote the length of a resident $r_i$'s individual preference list (regardless of whether $r_i$ is a single resident or $r_i$ belongs to a couple).

For a resident $r_i\in R$ (whether single or a member of a couple), let $pref(r_i, p)$ denote the hospital at position $p$ of $r_i$'s individual preference list.
%or on the projected preference list of coupled resident where $1\leq i\leq n_1$ and $1\leq p\leq l(r_i)$. %Further let $pref(h_j, q)$ denote the resident at position $q$ of $h_j$'s preference list where $1\leq j\leq n_2$ and $1\leq q\leq l(h_j)$.% 
%Let $pref( ( r_{2i}, r_{2i-1}), p)$ denote the hospital pair at position $p$ on the joint preference list of $(r_{2i-1}, r_{2i})$.
For an acceptable resident-hospital pair $(r_i, h_j)$, let $rank(h_j, r_i) = q$ denote the rank that hospital $h_j$ assigns resident $r_i$, where $1\leq q \leq l(h_j)$. Thus, $rank(h_j, r_i)$ is equal to the number of residents that $h_j$ prefers to $r_i$ plus 1.

Further, for each $j$ $(1\leq j\leq n_2)$ and $q$ $(1\leq q\leq l(h_j))$, let the set $R(h_j, q)$ contain resident-position pairs $(r_i,p)$ such that $r_i\in R$ is assigned a rank of $q$ by $h_j$ and $h_j$ is in position $p$ $(1\leq p\leq l(r_i))$ on $r_i$'s individual list.  Hence 
$$R(h_j, q) = \{(r_i, p)\in R \times \mathbb{Z} :  rank(h_j, r_i) = q \wedge 1\leq p\leq l(r_i)\wedge pref(r_i, p) = h_j\}.$$

\noindent \textbf{Variables in the IP model.} ~
For each $i$ $(1\leq i\leq n_1)$ and $p$ $(1\leq p\leq l(r_i))$, $J$ has a variable $x_{i,p}\in \{0,1\}$ such that $x_{i,p}=1$ if and only if $r_i$ is assigned to his $p^{th}$-choice hospital.  Also, for each $i$ ($1\leq i\leq n_1$) and $p=l(r_i)+1$, $J$ has a variable $x_{i,p}\in \{0,1\}$ such that $x_{i,p}=1$ if and only if $r_i$ is unassigned. Let $X = \{x_{i,p} : 1\leq i \leq n_1 \wedge 1\leq p \leq l(r_i) + 1\}$.

%Let $J$ be the IP model derived from $I$ as described in Section 3 of \cite{BMMcB14}. In this section we show how to modify $J$ to find a maximum cardinality most-stable\ matching in an instance of {\sc hrc}. We now describe the variables in the new model for {\sc min bp hrc} and give an example of how a stability constraint from the {\sc hrc} model presented in \cite{BMMcB14} can be adapted for the {\sc min bp hrc} context.

$J$ also contains variables $\theta_{i, p} \in \{0,1\}$ for each $i$ $(1\leq i\leq n_1)$ and $p$ $(1\leq p\leq l(r_i))$. The intuitive meaning of a variable $\theta_{i,p}$ is that $\theta_{i,p} = 1$ if and only if resident $r_i$ is involved in a blocking pair with the hospital at position $p$ on his individual preference list, either as a single resident or as part of a couple.

\medskip
\noindent \textbf{Constraints in the IP model.} ~  We firstly add constraints to $J$ which force every variable to be binary valued.  Next we ensure that matching constraints are satisfied, as follows.  As each resident $r_i\in R$ is assigned to exactly one hospital or is unassigned (but not both), $\sum_{p=1}^{l(r_i)+1} x_{i,p}=1$ must hold for all $i$ $(1\leq i\leq n_1)$.
%
%\begin{equation} \displaystyle \sum\limits_{p=1}^{l(r_i)+1} x_{i,p} %= 1 \end{equation}
%
Similarly, since a hospital $h_j$ may be assigned at most $c_j$ residents, $x_{i, p} = 1$ where $pref(r_i,p) = h_j$ for at most $c_j$ residents, and hence for all $j$ $(1\leq j\leq n_2)$, $\sum_{i=1}^{n_1}\sum_{p=1}^{l(r_i)} \{x_{i,p} \in X :  pref(r_i, p)=h_j\} \leq c_j$ must hold.
%
%\begin{equation} \displaystyle \sum\limits_{i=1}^{n_1} \sum\limits_{p=1}^{l(r_i)} \{x_{i,p} \in X :  pref(r_i, p)=h_j\} \leq c_j \end{equation}
%

For each couple $(r_{2i-1}, r_{2i})$, $r_{2i-1}$ is unassigned if and only if $r_{2i}$ is unassigned, and $r_{2i-1}$ is assigned to the hospital in position $p$ in their individual list if and only if $r_{2i}$ is assigned to the hospital in position $p$ in their individual list.  Thus for all $i$ $(1\leq i\leq c)$ and $p$ $(1\leq p\leq l(r_{2i-1})+1)$, $x_{2i-1,p} = x_{2i,p}$ must hold, 
%
%\begin{equation} \displaystyle x_{2i-1,p} = x_{2i,p} \end{equation}
%

The remaining constraints in $J$ allow the number of blocking pairs of a given matching to be counted.  Each such constraint deals with a specific type of blocking pair that satisfies a given part of Definition \ref{stability:MM}.  It allows a blocking pair to exist involving either (i) a single resident $r_i$ with the hospital at some position $p$ on his list, or (ii) a couple $(r_{2i-1},r_{2i})$ with the hospital pair at some position $p$ on their joint list, if and only if $\theta_{i,p}=1$.  We illustrate the construction of $J$ by giving the constraint corresponding to so-called ``Type 1'' blocking pairs, involving involve single residents, where Condition 1 of Definition \ref{stability:MM} is satisfied.  The other constraints may be dealt with in a similar fashion -- see Appendix \ref{section:IPModels_MINBPHRC} for further details.

\medskip\noindent
\textbf{Type 1 blocking pairs.}~ In a matching $M$ in $I$, if a single resident $r_i\in R$ is unassigned or has a worse partner than some hospital $h_j\in H$ where $pref(r_i, p)=h_j$ and $rank(h_j, r_i) =q$ then $h_j$ must be fully subscribed with better partners than $r_i$, for otherwise $(r_i,h_j)$ blocks $M$. Hence if $r_i$ is unassigned or has worse partner than $h_j$, i.e., $\sum\limits_{p^{\prime}=p+1}^{l(r_i)+1} x_{i,p^{\prime}}=1$, and $h_j$ is not fully subscribed with better partners than $r_i$, i.e.,  $\sum\limits_{q^{\prime}=1}^{q-1} \{x_{i^{\prime },p^{\prime \prime}} \in X :   ( r_{i^{\prime }}, p^{\prime \prime}) \in R(h_{j}, q^{\prime })\} < c_j$, then we require $\theta_{i,p}=1$ to count this blocking pair. Thus, for each $i$ $(2c+1\leq i\leq n_1)$ and $p$ $(1\leq p\leq l(r_i))$ we obtain the following constraint where $pref(r_i, p) = h_j$ and $rank(h_j, r_i)=q$:
%
%\begin{equation} \label{constraint:HRC2_5} \displaystyle c_j \sum\limits_{p^{\prime}=p+1}^{l(r_i)+1} x_{i,p^{\prime}} \leq \sum\limits_{q^{\prime}=1}^{q-1} \{x_{i^{\prime },p^{\prime \prime}} \in X :   rank(h_j,r_{i^{\prime }}) = q^{\prime } \wedge pref(r_{i^{\prime }}, p^{\prime \prime}) = h_j)\} \end{equation}
%
\[c_j \left(\left(\sum\limits_{p^{\prime}=p+1}^{l(r_i)+1} x_{i,p^{\prime}}\right) - \theta_{i,p}\right)  \leq \sum\limits_{q^{\prime}=1}^{q-1} \{x_{i^{\prime },p^{\prime \prime}} \in X :   ( r_{i^{\prime }}, p^{\prime \prime}) \in R(h_{j}, q^{\prime })\}.\]

\medskip
\noindent \textbf{Objective functions in the IP model.} ~ A maximum cardinality most-stable matching $M$ is a matching of maximum cardinality, taken over all most-stable matchings in $I$.  To compute a maximum most-stable matching in $J$, we apply two objective functions in sequence.

First we find an optimal solution in $J$ that minimises the number of blocking pairs. To this end we apply the objective function 
$\min \sum\limits_{i=1}^{n_1} \sum\limits_{p=1}^{l(r_i)} \theta_{i,p}$.
%shown in Equation \ref{constraint:MIN_BP_HRC_13} below.% A matching $M$ in $I$ with the minimum number of blocking pairs taken over all of the matchings in $I$ requires that the minimum number of $\theta_{i,p}$ must take the value of one. To minimise the sum over all of the values of $i$ and $p$ we apply the following objective function:
%\begin{equation} \label{constraint:MIN_BP_HRC_13} \displaystyle \min \sum\limits_{i=1}^{n_1} \sum\limits_{p=1}^{l(r_i)} \theta_{i,p} \end{equation}

The matching $M$ corresponding to an optimal solution in $J$ will be a most-stable\ matching in $I$. Let $k=|bp(M)|$.  Now we seek a maximum cardinality matching in $I$ with at most $k$ blocking pairs. Thus we add the following constraint to $J$, which ensures that, when maximising on cardinality, any solution also has at most $k$ blocking pairs: $\sum\limits_{i=1}^{n_1} \sum\limits_{p=1}^{l(r_i)} \theta_{i,p} \leq k$.
%\begin{equation} \label{constraint:MIN_BP_HRC_14} \displaystyle \sum\limits_{i=1}^{n_1} \sum\limits_{p=1}^{l(r_i)} \theta_{i,p} \leq k \end{equation}

The final step is to maximise the size of the matching, subject to the matching being most-stable.  This involves optimising for a second time, this time using the following objective function: $\max \sum\limits_{i=1}^{n_1} \sum\limits_{p=1}^{l(r_i)} x_{i,p}$.
%A maximum cardinality most-stable\ matching $M$ is a matching in which the maximum number of residents are assigned in $M$ subject to having the minimum possible number of blocking pairs taken over all of the matchings admitted by $I$. To maximise the size of the matching found, subject to Constraint \eqref{constraint:MIN_BP_HRC_14} holding, we also apply the following objective function:
%\begin{equation} \label{constraint:MIN_BP_HRC_15} \displaystyle \max \sum\limits_{i=1}^{n_1} \sum\limits_{p=1}^{l(r_i)} x_{i,p}.\end{equation}

The following result, which establishes the correctness of the IP formulation, is proved in Appendix \ref{section:IPModels_MINBPHRC}. 
\begin{theorem1}
Given an instance $I$ of {\sc min bp hrc}, let $J$ be the corresponding IP model as defined above. A maximum cardinality most-stable matching in $I$ is exactly equivalent to an optimal solution to $J$.
\end{theorem1}

We remark that the IP model presented in this section develops the earlier model for {\sc hrc} \cite{BMMcB14} with the addition of the $\theta_{i,p}$ variables. There are similarities between the constraints (with these variables omitted) when comparing the two models.  However in the {\sc hrc} model \cite{BMMcB14} essentially all stability constraints had to be satisfied, whereas in the {\sc min bp hrc} model a blocking pair is allowed at the expense of a $\theta_{i,p}$ variable having value 1, which allows the number of blocking pairs to be counted.  Suitable placement of the $\theta_{i,p}$ variables within the constraints from the {\sc hrc} model allows this condition on the $\theta_{i,p}$ variables to be enforced.

\section{Empirical results from the IP model for {\sc min bp hrc}}
\label{section:MIN_BP_HRC_IPexperiments}

In this section we present data from an empirical evaluation of an implementation of the IP model for finding a maximum cardinality most-stable matching in an instance of {\sc min bp hrc}. We considered the following properties for randomly-generated {\sc hrc} instances: the time taken to find a maximum cardinality most-stable matching, the size of a maximum cardinality most-stable matching and the number of blocking pairs admitted by a most-stable matching. We show how these properties varied as we modified the number of residents, the percentage of residents involved in couples, the number of hospitals and the lengths of residents' preference lists in the constructed instances.

\medskip

\noindent \textbf{Methodology.} ~ We ran all the experiments on an implementation of the IP model using the CPLEX 12.4 Java Concert API applied to randomly-generated instances of {\sc hrc}\footnote{All generated instances can be obtained from \url{http://dx.doi.org/10.5525/gla.researchdata.303}.}.  In these instances, the preference lists of residents and hospitals were constructed to take into account of the fact that, in reality, some hospitals and residents are more popular than others, respectively.  Typically, the most popular hospital in the SFAS context had 5-6 times as many applicants as the least popular, and the numbers of applicants to the other hospitals were fairly uniformly distributed between the two extremes.  Our constructed instances reflected this real-world behaviour.  For more details about the construction of the instances and the correctness testing methodology, the reader is referred to \cite[Chapters 6,7]{McB15}.  

All experiments were carried out on a desktop PC with an Intel i5-2400 3.1Ghz processor with 8Gb of memory running Windows 7. To find a most-stable matching in an instance $I$ of {\sc hrc} we applied the following procedure. We first used the {\sc hrc} IP implementation presented in \cite{BMMcB14} to find a maximum cardinality stable matching $M$ in $I$ if one exists. Clearly, if $I$ is solvable then $M$ is a maximum cardinality most-stable matching. However, if $I$ was found to be unsolvable, we applied the {\sc min bp hrc} IP model to $I$. In this case we applied a lower bound of 1 to the number of blocking pairs in a most-stable matching in $I$ since we knew that no stable matching existed.  All instances were allowed to run to completion.  We remark that the {\sc min bp hrc} model appears to be much more difficult to solve than the {\sc hrc} model presented in \cite{BMMcB14}, and thus the largest instances sizes considered here are smaller than the largest ones generated in the experimental evaluation in \cite{BMMcB14}.

%The mean time taken to find a maximum cardinality most-stable matching taken over those instances admitting a stable matching, the mean time taken to find a maximum cardinality most-stable matching taken over those instances that did not admit a stable matching, the mean size of the maximum cardinality most-stable matching taken over those instances admitting a stable matching, the mean size of a maximum cardinality most-stable matching taken over only those instances that did not admit a stable matching and then mean and maximum number of blocking pairs admitted by the instances are shown in the data that follows.

\begin{figure}[t!]
\includegraphics[width=\textwidth, height=4.55cm]{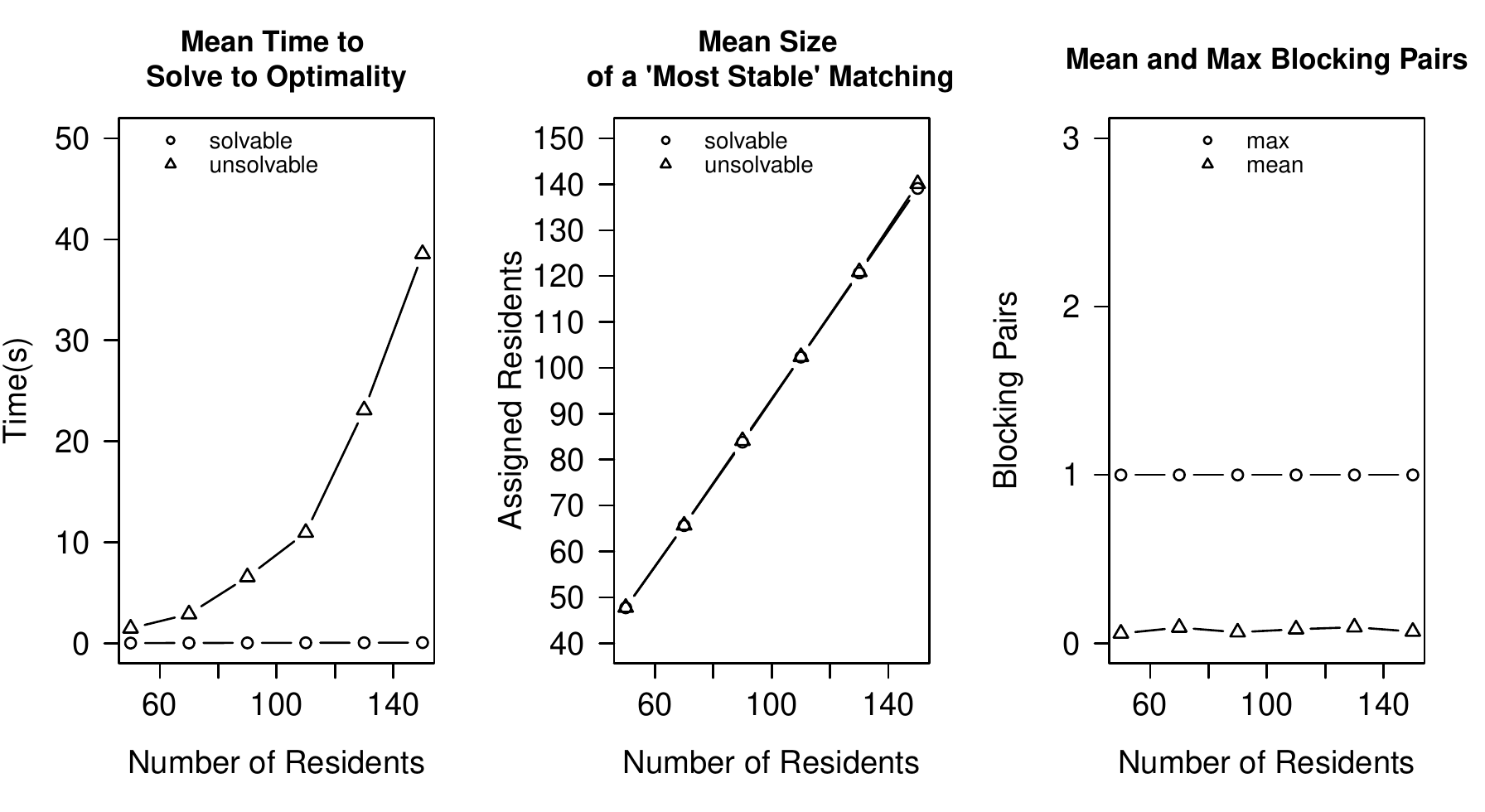}
\vspace{-5mm}
\caption{Empirical results for Experiment 1.}
\label{chart:min_bp_hrc_experiment1_data}
\end{figure}

\medskip
\noindent \textbf{Experiment 1.} ~ In the first experiment we increased the number of residents while maintaining a constant ratio of couples, hospitals and posts to residents. For various values of $x$ $(50 \leq x \leq 150)$ in increments of $20$, $1000$ randomly generated instances were created containing $x$ residents, $0.1x$ couples (and hence $0.8x$ single residents) and $0.1x$ hospitals with $x$ available posts that were randomly distributed amongst the hospitals. Each resident's preference list contained a minimum of 3 and a maximum of 5 hospitals. Figure \ref{chart:min_bp_hrc_experiment1_data} (and indeed all the figures in this section) shows the mean time taken to find a maximum cardinality most-stable matching, the mean size of a maximum cardinality most-stable solution (in each case over both solvable and unsolvable instances), and the mean and maximum number of blocking pairs admitted by most-stable matchings.

The results show that the time taken to find an optimal solution increases with $x$, with the {\sc min bp hrc} formulation being more difficult to solve in general than the {\sc hrc} formulation.  The mean size of an optimal solution increases with $x$ for both solvable and unsolvable instances (it is around 95\% of $x$ for $x=50$, decreasing to around 93\% of $x$ for $x=150$, with the optimal matching size for unsolvable instances being very slightly larger than that for solvable instances).  Perhaps most interestingly, the maximum number of blocking pairs was 1, with the mean at most 0.1, and the mean number of unsolvable instances being 77.
%data in Figure \ref{chart:min_bp_hrc_experiment1_data} show that the mean time taken to find a maximum cardinality most-stable matching tends to increase as we increase the number of residents in the instances. The increase is more pronounced for those instances not admitting a stable matching. Instances admitting a stable matching are solved by the {\sc hrc} IP model and are not considered by the {\sc min bp hrc} model. The instances not admitting a stable matching must be solved by the {\sc min bp hrc} IP model and this model takes much longer to find an optimal solution than the corresponding {\sc hrc} IP model. 

%The data in Figure \ref{chart:min_bp_hrc_experiment1_data} show that the mean size of a maximum cardinality most-stable matching for those instances not admitting a stable matching is greater in all cases than the mean size of a maximum cardinality most-stable matching for those instances that admitting a stable matching. We conjecture that the set of matchings admitting exactly zero blocking pairs is likely to be smaller than the set of matchings admitting more than zero blocking pairs. Hence the latter set is more likely to admit a matching with a larger maximum cardinality most-stable matching. The data in Figure \ref{chart:min_bp_hrc_experiment1_data} show that the maximum number of blocking pairs in a maximum cardinality most-stable matching is exactly one for all of the instance sizes considered. The mean number of blocking pairs in a maximum cardinality most-stable matching does not appear to alter significantly with the size of the instance. 
\medskip

\begin{figure}[t]
\includegraphics[width=\textwidth, height=4.55cm]{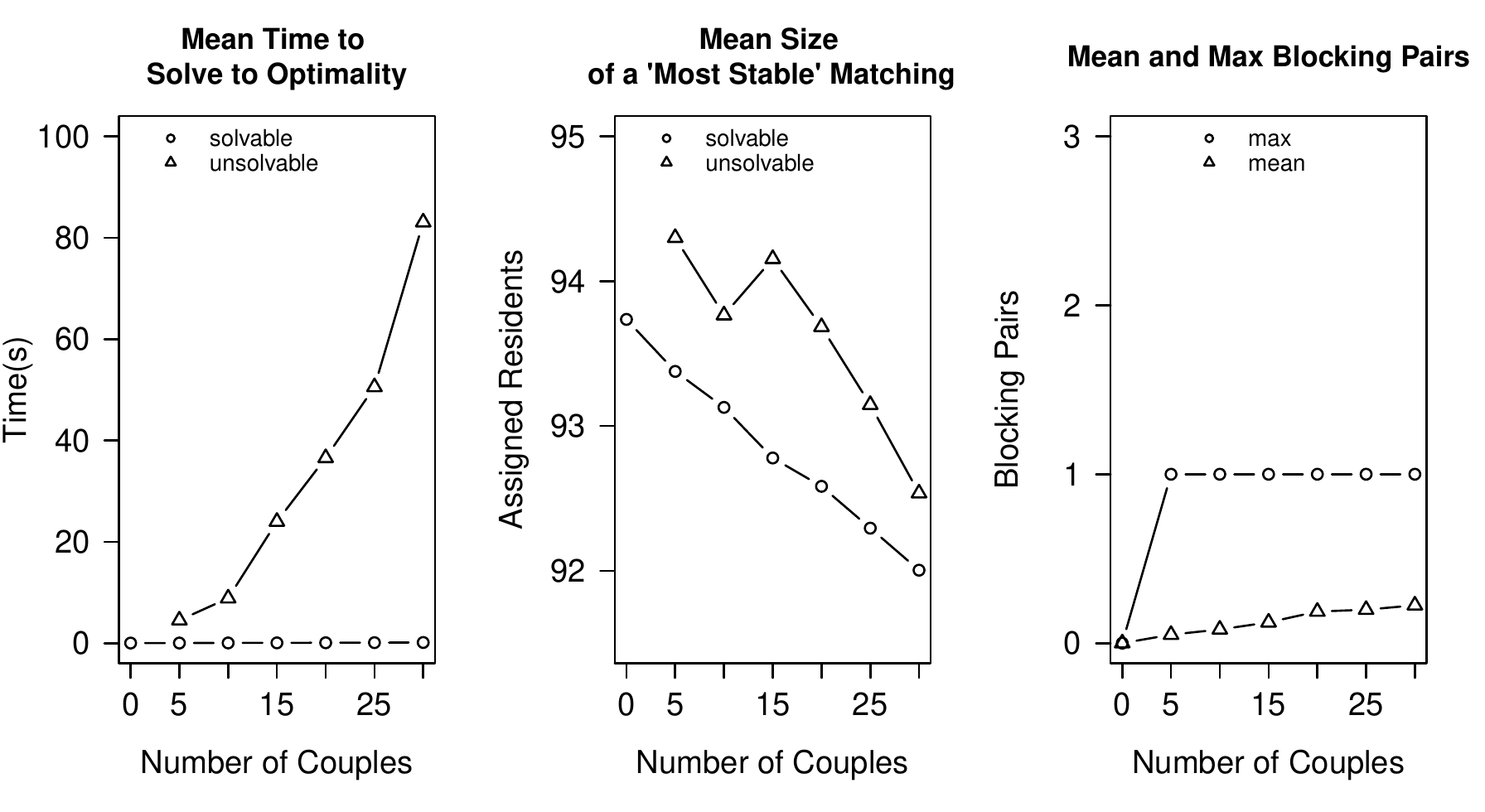}
\vspace{-5mm}
\caption{Empirical results for Experiment 2.}
\label{chart:min_bp_hrc_experiment2_data}
\end{figure}

\noindent \textbf{Experiment 2.} ~ In our second experiment we increased the percentage of residents involved in couples while maintaining the same numbers of residents, hospitals and posts. For various values of $x$ $(0 \le x \le 30)$ in increments of $5$, $1000$ randomly generated instances were created containing $100$ residents, $x$ couples (and hence $100-2x$ single residents) and $10$ hospitals with $100$ available posts that were unevenly distributed amongst the hospitals. Each resident's preference list contained a minimum of 3 and a maximum of 5 hospitals. The results for all values of $x$ are displayed in Figure \ref{chart:min_bp_hrc_experiment2_data}.

The results show that the time taken to find an optimal solution increases with $x$; again the {\sc min bp hrc} formulation is more difficult to solve in general than the {\sc hrc} formulation.  The mean size of an optimal solution decreases with $x$ for both solvable and unsolvable instances; again the optimal matching size for unsolvable instances is slightly larger than that for solvable instances.  As for Experiment 1, the maximum number of blocking pairs was 1, with the number of unsolvable instances increasing from 50 for $x=5$ to 224 for $x=30$.

\begin{figure}[t]
\includegraphics[width=\textwidth, height=4.55cm]{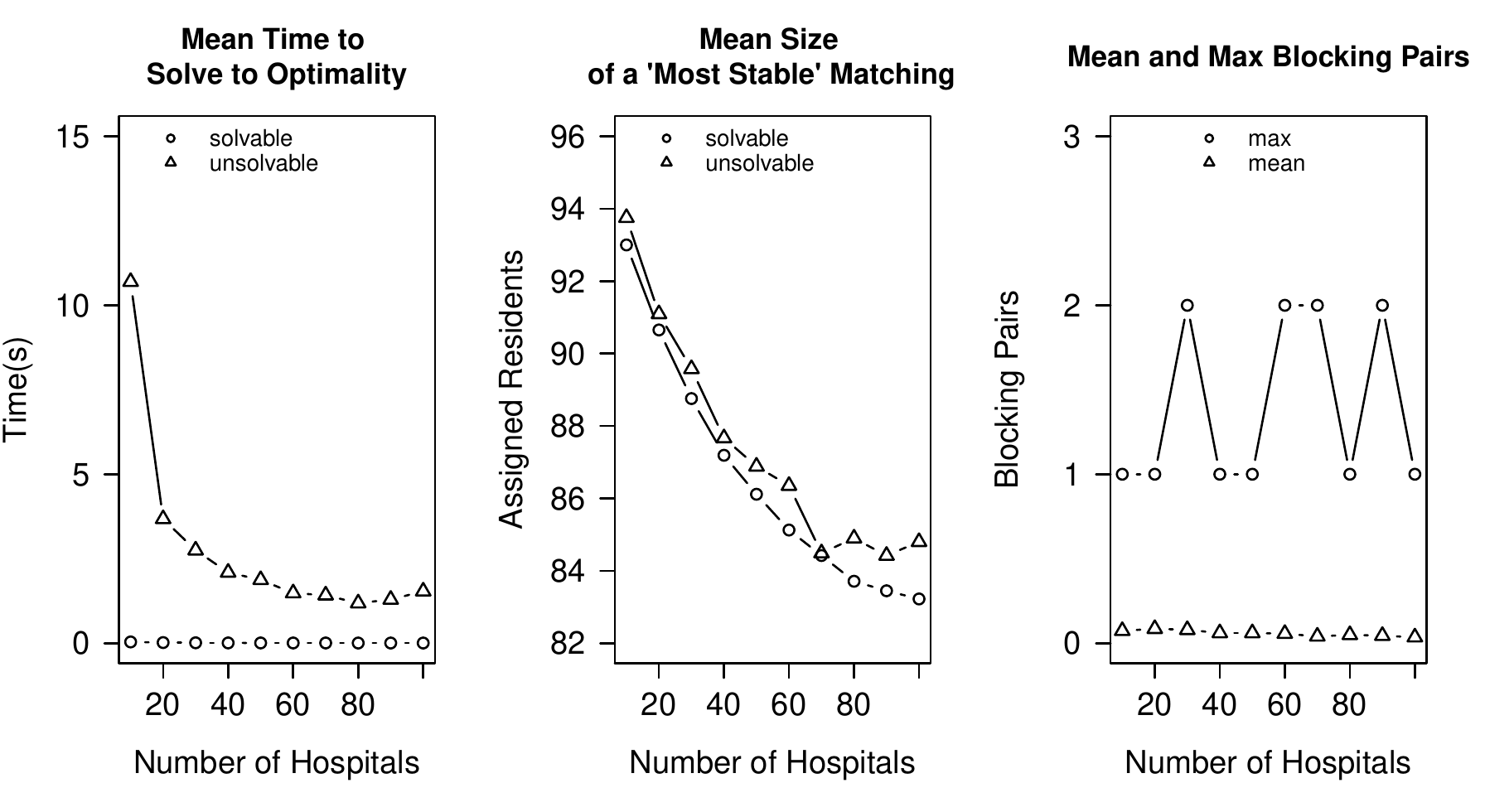}
\vspace{-5mm}
\caption{Empirical results for Experiment 3.}
\label{chart:min_bp_hrc_experiment3_data}
\end{figure}

\medskip
\noindent \textbf{Experiment 3.} In our third experiment we increased the number of hospitals in the instance while maintaining the same numbers of residents, couples and posts. For various values of $x$ $(10 \le x \le 100)$ in increments of $10$, $1000$ randomly generated instances were created containing $100$ residents, $10$ couples (and hence $80$ single residents) and $x$ hospitals with $100$ available posts that were unevenly distributed amongst the hospitals. Each resident's preference list contained a minimum of 3 and a maximum of 5 hospitals. The results for all values of $x$ are displayed in Figure \ref{chart:min_bp_hrc_experiment3_data}.

The results show that the time taken to find an optimal solution decreases with $x$; again the {\sc min bp hrc} model solution time is slower than that for the {\sc hrc} model.  Clearly the problem is becoming less constrained as the number of hospitals increases.  Also the mean size of an optimal solution decreases with $x$ for both solvable and unsolvable instances; again the optimal matching size for unsolvable instances is slightly larger than that for solvable instances.  This time the maximum number of blocking pairs was 2, with the mean number of blocking pairs decreasing from $0.08$ for $x=20$ to $0.04$ for $x=100$.

\begin{figure}[t]
\includegraphics[width=\textwidth, height=4.55cm]{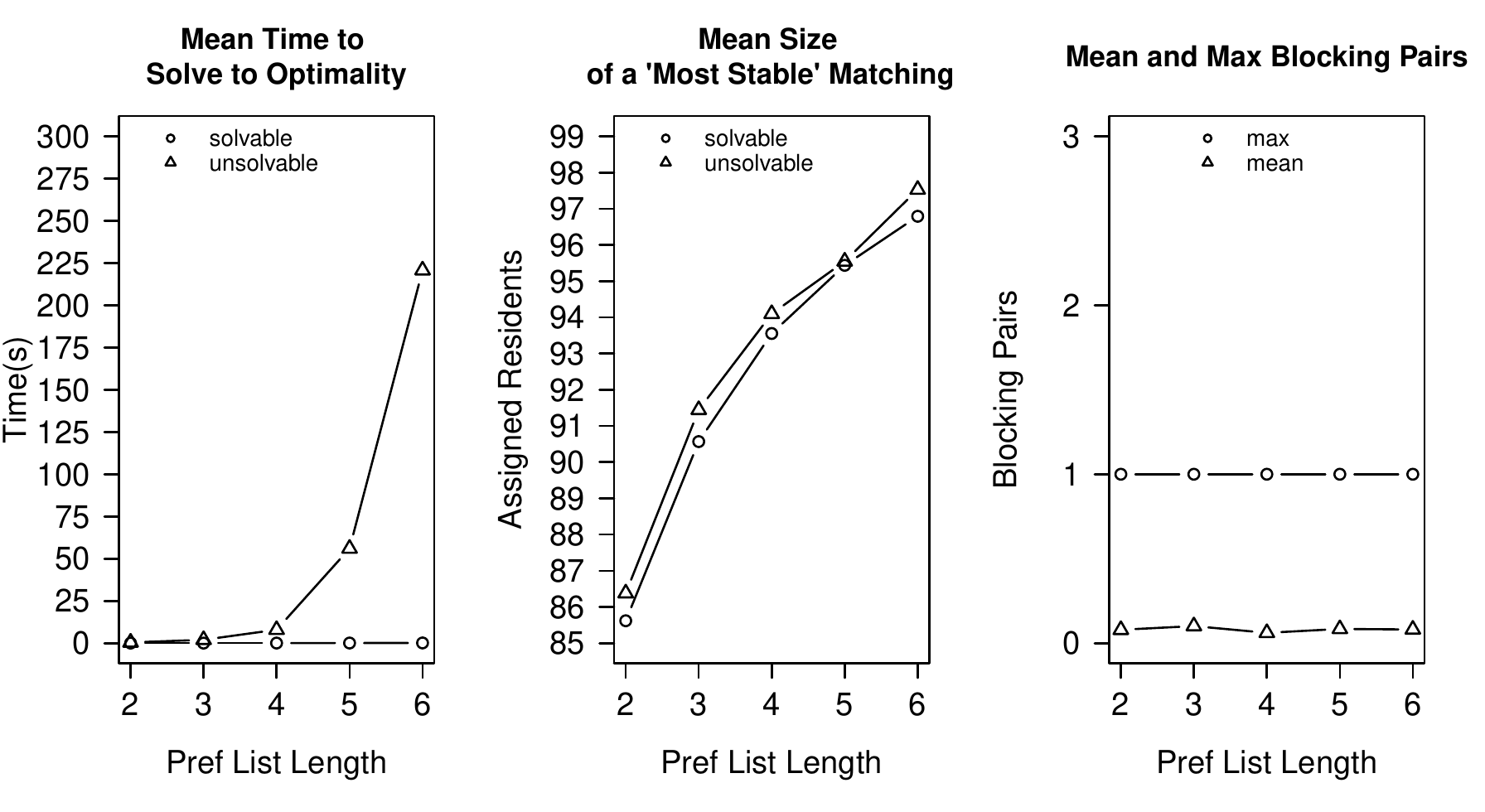}
\vspace{-5mm}
\caption{Empirical results for Experiment 4.}
\label{chart:min_bp_hrc_experiment4_data}
\end{figure}

\medskip
\noindent \textbf{Experiment 4.} ~ In our last experiment, we increased the length of the individual preference lists for the residents in the instance while maintaining the same numbers of residents, couples, hospitals and posts. For various values of $x$ $(2 \le x \le 6)$, $1000$ randomly generated instances were created containing $100$ residents, $10$ couples (and hence $80$ single residents) and $10$ hospitals with $100$ available posts that were unevenly distributed amongst the hospitals. Each resident's preference list contained exactly $x$ hospitals. The results for all values of $x$ are displayed in Figure \ref{chart:min_bp_hrc_experiment4_data}.

The results show that increasing the preference list length makes the problem harder to solve; again the {\sc min bp hrc} model is slower to solve than the  {\sc hrc} model.  Also the mean size of an optimal solution increases with $x$ for both solvable and unsolvable instances as more options become available in the preference lists (from 86.4 for $x=2$ to 97.5 for $x=6$ in the case of unsolvable instances); again the optimal matching size for unsolvable instances is slightly larger than that for solvable instances.  The maximum number of blocking pairs was 1, with the mean at most 0.1, and the mean number of unsolvable instances being 81.

\medskip
\noindent \textbf{Discussion.} ~ The results presented in this section suggest that, even as we increase the number of residents or hospitals, the percentage of residents involved in a couple or the length of the residents' preference lists, the number of blocking pairs admitted by a most-stable matching is very low. For most of the 28,000 instances generated in our experimental evaluation, the most-stable matchings found admitted at most 1 blocking pair, and the maximum number of blocking pairs admitted by any most-stable matching was never more than 2. These findings are essentially consistent with the results of Nguyen and Vohra \cite{NV14}, who showed that an unsolvable {\sc hrc} instance only requires a small amount of perturbation in order to become solvable.  Further empirical investigation is required to determine whether this behaviour is replicated for larger {\sc hrc} instance sizes.

\section{A Constraint Programming model for {\sc min bp hrc}}
\label{section:CPmodel}
In addition to the IP model, we designed a Constraint Programming model
for {\sc min bp hrc} and implemented this using the MiniZinc constraint modelling
language.

We assume that residents' preference lists are given by integer variables \emph{rpref}$[i][j]$, which play a similar role to \emph{pref}$(r_i,j)$ in the IP model, and that hospitals' ranking arrays are given by integer variables \emph{hrank}$[h,i]$, which are analogous to \emph{rank}$(h_j,r_i)$ in the IP model.  The lengths of the preference lists of a resident $r_i$ and a hospital $h_j$ are given by \emph{rpref}\_\emph{len}$[i]$ and \emph{hpref}\_\emph{len}$[j]$ respectively.  The capacity of a hospital $h_j$ is given by $hosp\_cap[j]$.

For each single resident $r_i$, the model includes an integer variable
\emph{single}\_\emph{pos}$[i]$ with domain $(1,\dots,l(r_i)+1)$, where $l(r_i)$ is the value of \emph{rpref}\_\emph{len}$[i]$, which takes the value $j$ if $r_i$ is
assigned her $j$th-choice hospital, or $l(r_i)+1$ if $r_i$ is unassigned. For
each couple $i$, we include an integer variable $coup\_pos[i]$ with a similar
interpretation.

Each single resident's \emph{single}\_\emph{pos}$[i]$ variable is channelled to an array of
$l(r_i)$ boolean variables \emph{single}\_\emph{assigned}$[i]$, such that \emph{single}\_\emph{assigned}$[i][j]=$ {\sf true} 
if and only if \emph{single}\_\emph{pos}$[i]=j$, and a variable \emph{single}\_\emph{unassigned}$[i]$,
such that \emph{single}\_\emph{unassigned}$[i]=$ {\sf true} if and only if \emph{single}\_\emph{pos}$[i]=l(r_i) +
1$. Similarly, we have boolean \emph{coup}\_\emph{assigned} and \emph{coup}\_\emph{unassigned} variables for
each couple.

For each hospital $i$, and each position $j$ on hospital $i$'s preference list, we
have a boolean variable \emph{hosp}\_\emph{assigned}$[i][j]$ which is {\sf true} if and only if
hospital $i$ is assigned its $j$th-choice resident. We include a constraint to
ensure that \emph{hosp}\_\emph{assigned}$[i][j]=$ {\sf true} if and only if a corresponding
\emph{single}\_\emph{assigned} or \emph{coup}\_\emph{assigned} variable is also {\sf true}.  
Furthermore, each hospital has a linear inequality constraint to ensure that its capacity is not exceeded.

For each position on the preference list of a single resident or couple, we
create a boolean variable \emph{single}\_\emph{bp}$[i][j]$ or \emph{coup}\_\emph{bp}$[i][j]$ indicating whether the
resident or couple, along with their $j$th-choice hospital, constitutes
a blocking pair. For each type of blocking pair, we define a set of constraints and then give some brief intuition.
%As an example, we show here the constraints for Type 1 blocking pairs.

\medskip
\noindent {\bf Type 1 blocking pairs}
\begin{tabbing}
and \= and \= and \= and \= and \= \kill
{\bf constraint } forall ($i$ {\bf in} \emph{Singles}) (\\
  \> {\bf forall}($j$ {\bf in} $1\dots$\emph{rpref}\_\emph{len}$[i]$) (\\
  \> \> let \{{\bf int:} $h=$\emph{rpref}$[i,j]$, {\bf int:} $q=$\emph{hrank}$[h,i]$\} {\bf in} \\
  \> \> \emph{single}\_\emph{pos}$[i] > j$ $\wedge$ \emph{hosp}\_\emph{would}\_\emph{prefer}$(h,q) \Rightarrow$ \emph{single}\_\emph{bp}$[i,j]$ ));
%  \> )\\
%);
\end{tabbing}

The \emph{hosp}\_\emph{would}\_\emph{prefer} predicate for a hospital $h$ and a position $q$ on the preference list of $h$ takes the value {\sf true} if and only if $h$ has fewer than \emph{hosp}\_\emph{cap}$[h]$ assigned residents in positions strictly preferable to position $q$ on its preference list.  (Note the redundancy in this predicate: all we actually need is the first $sum(\dots)(\dots)<$ \emph{hosp}\_\emph{cap}$[h]$ constraint; the $sum(\dots)(\dots)>0$ constraint improves propagation.)

\begin{tabbing}
and \= and \= and \= and \= and \= \kill
{\bf predicate} \emph{hosp}\_\emph{would}\_\emph{prefer}({\bf int}:$h$, {\bf int}:$q$) =\\
  \> {\bf if} $q\leq hosp\_cap[h]$ {\bf then}\\
  \> \> {\sf true}\\
  \> {\bf else}\\
  \> \> {\bf sum}($k$ {\bf in} $1\dots q-1$)({\bf bool2int}(\emph{hosp}\_\emph{assigned}$[h,k]$)) $<$ \emph{hosp}\_\emph{cap}$[h]$ $\vee$\\
  \> \> {\bf sum}($k$ {\bf in} $q+1\dots$\emph{hpref}\_\emph{len}$[h]$)({\bf bool2int}(\emph{hosp}\_\emph{assigned}$[h,k]$)) $>$ 0\\
  \> {\bf endif};
\end{tabbing}

The constraint for Type 1 blocking pairs thus sets \emph{single}\_\emph{bp}$[i,j]$ to {\sf true} if and only if $r_i$ is unassigned or prefers $h$ to his partner, and $h$ is undersubscribed or prefers $r_i$ at least one of its assignees, where $h=$ \emph{rpref}$[i,j]$.  
%The remaining constraints that make up the CP model are shown in Appendix \ref{sec:full-cp-model}.
%\section{A Constraint Programming model for {\sc min bp hrc}}
%\label{sec:full-cp-model}
%In Section \ref{section:CPmodel}, we presented the constraints that correspond to the identification of Type 1 blocking pairs.  We now present the constraints corresponding to the remaining blocking pair types from Definition \ref{stability:MM}.

\medskip
\noindent {\bf Type 2a/b blocking pairs}
\begin{tabbing}
and \= and \= and \= and \= and \= \kill
{\bf constraint } forall ($i$ {\bf in} \emph{Couples}) (\\
  \> let \{{\bf int:} $r1=$ \emph{first}\_\emph{in}\_\emph{couple}$(i)$, {\bf int:} $r2=$ \emph{second}\_\emph{in}\_\emph{couple}$(i)$\} {\bf in} \\
  \> {\bf forall}($j$ {\bf in} $1\dots$\emph{rpref}\_\emph{len}$[r1]$ (\\
  \> \> let \{{\bf int:} $h1=$ \emph{rpref}$[r1,j]$, {\bf int:} $h2=$ \emph{rpref}$[r2,j]$, \\
  \> \> \> {\bf int:} $q1= $\emph{hrank}$[h1,r1]$, {\bf int:} $q2=$ \emph{hrank}$[h2,r2]$\} {\bf in} \\
  \> \> \emph{coup}\_\emph{pos}$[i] > j$ $\wedge$ \\
  \> \> ((\emph{hosp}\_\emph{would}\_\emph{prefer}\_\emph{exc}\_\emph{partner}$(h1,h2,q1,q2)$ $\wedge$ \\
  \> \> \> $h2 = $ \emph{rpref}$[r2,$\emph{coup}\_\emph{pos}$[i]])$ $\vee$ \\
  \> \> (\emph{hosp}\_\emph{would}\_\emph{prefer}\_\emph{exc}\_\emph{partner}$(h2,h1,q2,q1)$ $\wedge$ \\
  \> \> \> $h1 = $ \emph{rpref}$[r1,$\emph{coup}\_\emph{pos}$[i]])$) \\
  \> \> $\Rightarrow$ \emph{coup}\_\emph{bp}$[i,j]$\\ 
  \> )\\
);
\end{tabbing}

The \emph{hosp}\_\emph{would}\_\emph{prefer}\_\emph{exc}\_\emph{partner} predicate on inputs $h1$, $h2$, $q1$, $q2$ (where $h1$, $h2$ are hospitals and $q1$, $q2$ are positions on their preference lists respectively) takes the value {\sf true} if and only if (a) $h1=h2$, $q1<q2$ and the number of $h1$'s assignees that it prefers to its $q1$th choice is less than $hosp\_cap[h1]-1$, or (b) $h1\neq h2$ or $q1>q2$ and the number of $h1$'s assignees that it prefers to its $q1$th choice is less than $hosp\_cap[h1]$.
\begin{tabbing}
and \= and \= and \= and \= and \= \kill
{\bf predicate} \emph{hosp}\_\emph{would}\_\emph{prefer}\_\emph{exc}\_\emph{partner}$$({\bf int}:$h1$, {\bf int}:$h2$, {\bf int}:$q1$, {\bf int}:$q2$) =\\
  \> {\bf if} $h1=h2$ $\wedge$ $q1<q2$ {\bf then}\\
  \> \> {\bf sum}($k$ {\bf in} $1\dots q1-1$)({\bf bool2int}(\emph{hosp}\_\emph{assigned}$[h1,k]$)) $<$ \emph{hosp}\_\emph{cap}$[h1]-1$ \\
  \> {\bf else}\\
  \> \> {\bf sum}($k$ {\bf in} $1\dots q1-1$)({\bf bool2int}(\emph{hosp}\_\emph{assigned}$[h1,k]$)) $<$ \emph{hosp}\_\emph{cap}$[h1]$ \\
  \> {\bf endif};
\end{tabbing}

The constraint for Type 2a/b blocking pairs thus sets \emph{coup}\_\emph{bp}$[i,j]$ to {\sf true} if and only if couple $(r1,r2)$ prefer hospital pair $(h1,h2)$ to their joint assignment $(h3,h4)$, where \emph{either}
\begin{enumerate}
\item[(a)] $h2=h4$ and either $h1$ is undersubscribed or prefers $r1$ to at least one assignee that is not $r2$ (if $r2$ is assigned to $h1$) \emph{or}
\item[(b)] $h1=h3$ and either $h2$ is undersubscribed or prefers $r2$ to at least one assignee that is not $r1$ (if $r1$ is assigned to $h2$).
\end{enumerate}

\medskip
\noindent {\bf Type 3a blocking pairs}
\begin{tabbing}
and \= and \= and \= and \= and \= \kill
{\bf constraint } forall ($i$ {\bf in} \emph{Couples}) (\\
  \> let \{{\bf int:} $r1=$ \emph{first}\_\emph{in}\_\emph{couple}$(i)$, {\bf int:} $r2=$ \emph{second}\_\emph{in}\_\emph{couple}$(i)$\} {\bf in} \\
  \> {\bf forall}($j$ {\bf in} $1\dots$ \emph{rpref}\_\emph{len}$[r1]$ {\bf where} \emph{rpref}$[r1,j]$ != \emph{rpref}$[r2,j])$ (\\
  \> \> let \{{\bf int:} $h1= $\emph{rpref}$[r1,j]$, {\bf int:} $h2=$ \emph{rpref}$[r2,j]$, \\
  \> \> \> {\bf int:} $q1=$ \emph{hrank}$[h1,r1]$, {\bf int:} $q2=$ \emph{hrank}$[h2,r2]$\} {\bf in} \\
  \> \> \emph{hosp}\_\emph{would}\_\emph{prefer}$(h1,q1)$ $\wedge$ \emph{hosp}\_\emph{would}\_\emph{prefer}$(h2,q2)$ $\wedge$ \\
  \> \> $h1$ != \emph{rpref}$[r1,$\emph{coup}\_\emph{pos}$[i]]$ $\wedge$ $h2$ != \emph{rpref}$[r2,$\emph{coup}\_\emph{pos}$[i]]$ $\wedge$ \\
  \> \> \emph{coup}\_\emph{pos}$[i]> j$ $\Rightarrow$ \emph{coup}\_\emph{bp}$[i,j]$\\ 
  \> )\\
);
\end{tabbing}

The constraint for Type 3a blocking pairs thus sets \emph{coup}\_\emph{bp}$[i,j]$ to {\sf true} if and only if couple $(r1,r2)$ are unassigned or prefer $(h1,h2)$ to their joint assignment, whilst for each $k\in \{1,2\}$, $hk$ is undersubscribed or prefers $rk$ to at least one of its assignees, where $(r1,r2)$ is the $i$th couple and $(h1,h2)$ is the hospital pair at position $j$ of their joint list.

\medskip
\noindent {\bf Type 3b/c/d blocking pairs}
\begin{tabbing}
and \= and \= and \= and \= and \= \kill
{\bf constraint } forall ($i$ {\bf in} \emph{Couples}) (\\
  \> let \{{\bf int:} $r1=$ \emph{first}\_\emph{in}\_\emph{couple}$(i)$, {\bf int:} $r2=$ \emph{second}\_\emph{in}\_\emph{couple}$(i)$\} {\bf in} \\
  \> {\bf forall}($j$ {\bf in} $1\dots$ \emph{rpref}\_\emph{len}$[r1]$ {\bf where} \emph{rpref}$[r1,j]$ = \emph{rpref}$[r2,j])$ (\\
  \> \> let \{{\bf int:} $h=$ \emph{rpref}$[r1,j]$, {\bf int:} $q1=$ \emph{hrank}$[h,r1]$, \\
  \> \> \> {\bf int:} $q2=$ \emph{hrank}$[h,r2]$\} {\bf in} \\
  \> \> {\bf if} $q1 < q2$ {\bf then}\\
  \> \> \> \emph{hosp}\_\emph{would}\_\emph{prefer}$2(h,q1)$ $\wedge$ \emph{hosp}\_\emph{would}\_\emph{prefer}$(h,q2)$ \\
  \> \> {\bf else}\\
  \> \> \> \emph{hosp}\_\emph{would}\_\emph{prefer}$(h,q1)$ $\wedge$ \emph{hosp}\_\emph{would}\_\emph{prefer}$2(h,q2)$ \\
  \> \> {\bf end if} $\wedge$ \\
  \> \> \emph{coup}\_\emph{pos}$[i]> j$ $\wedge$ \\
  \> \> $h$ $!=$ \emph{rpref}$[r1,$\emph{coup}\_\emph{pos}$[i]]$ $\wedge$ $h$ $!=$ \emph{rpref}$[r2,$\emph{coup}\_\emph{pos}$[i]]$ \\
  \> \> $\Rightarrow$ \emph{coup}\_\emph{bp}$[i,j]$\\ 
  \> )\\
);
\end{tabbing}

The \emph{hosp}\_\emph{would}\_\emph{prefer}$2$ predicate for a hospital $h$ and a position $q$ on the preference list of $h$ takes the value {\sf true} if and only if $h$ has fewer than \emph{hosp}\_\emph{cap}$[h]-1$ assigned residents in positions strictly preferable to position $q$ on its preference list.  (Note the redundancy in this predicate: all we actually need is the first $sum(\dots)(\dots)<$ \emph{hosp}\_\emph{cap}$[h] - 1$ constraint; the $sum(\dots)(\dots)>1$ constraint improves propagation.)

\begin{tabbing}
and \= and \= and \= and \= and \= \kill
{\bf predicate} \emph{hosp}\_\emph{would}\_\emph{prefer}$2$({\bf int}:$h$, {\bf int}:$q$) =\\
  \> {\bf sum}($k$ {\bf in} $1\dots q-1$)({\bf bool2int}(\emph{hosp}\_\emph{assigned}$[h,k]$)) $<$ $hosp\_cap[h] - 1$ $\vee$\\
  \> {\bf sum}($k$ {\bf in} $q+1\dots$ \emph{hpref}\_\emph{len}$[h]$)({\bf bool2int}(\emph{hosp}\_\emph{assigned}$[h,k]$)) $>$ 1 ;
\end{tabbing}

The constraint for Type 3b/c/d blocking pairs thus sets \emph{coup}\_\emph{bp}$[i,j]$ to {\sf true} if and only if couple $(r1,r2)$ are unassigned or prefer $(h,h)$ to their joint assignment, whilst $h$ either has two free posts (Type 3b), or $h$ has one free post and prefers one of $r1$ or $r2$ to at least one of its assignees (Type 3c), or $h$ is full and and prefers $r1$ to some assignee $rk$, and prefers $r2$ to at least one of its assignees apart from $rk$ (Type 3d), where $(r1,r2)$ is the $i$th couple and $(h,h)$ is the hospital pair at position $j$ of their joint list.

\medskip \noindent
{\bf Experiments.} The CP model was solved using the lazy clause solver Chuffed \cite{ZZZ10} on the same machine that was used for the experiments on the IP model as reported in Section \ref{section:MIN_BP_HRC_IPexperiments}.  All instances were allowed to run to completion.  We present results on the runtime of the CP model both with and without presolving.  The presolve step, when included, specifies in advance which set $S$ of resident-hospital pairs will block the solution (in practice we try out values of $k=0,1,2,\dots$ and generate all subsets $S$ of size $k$ until we reach a feasible solution) and then performs preference list deletions in the knowledge that the pairs in $S$ will block.  This allows large reductions in the model size, and works well because the number of blocking pairs admitted by a most-stable matching is generally very small, as we saw in Section \ref{section:MIN_BP_HRC_IPexperiments}. We did not use presolve with the IP model, but we note that it may be possible to solve the IP model more quickly by carrying out this step.

\begin{figure}[t!]
\includegraphics[width=\textwidth]{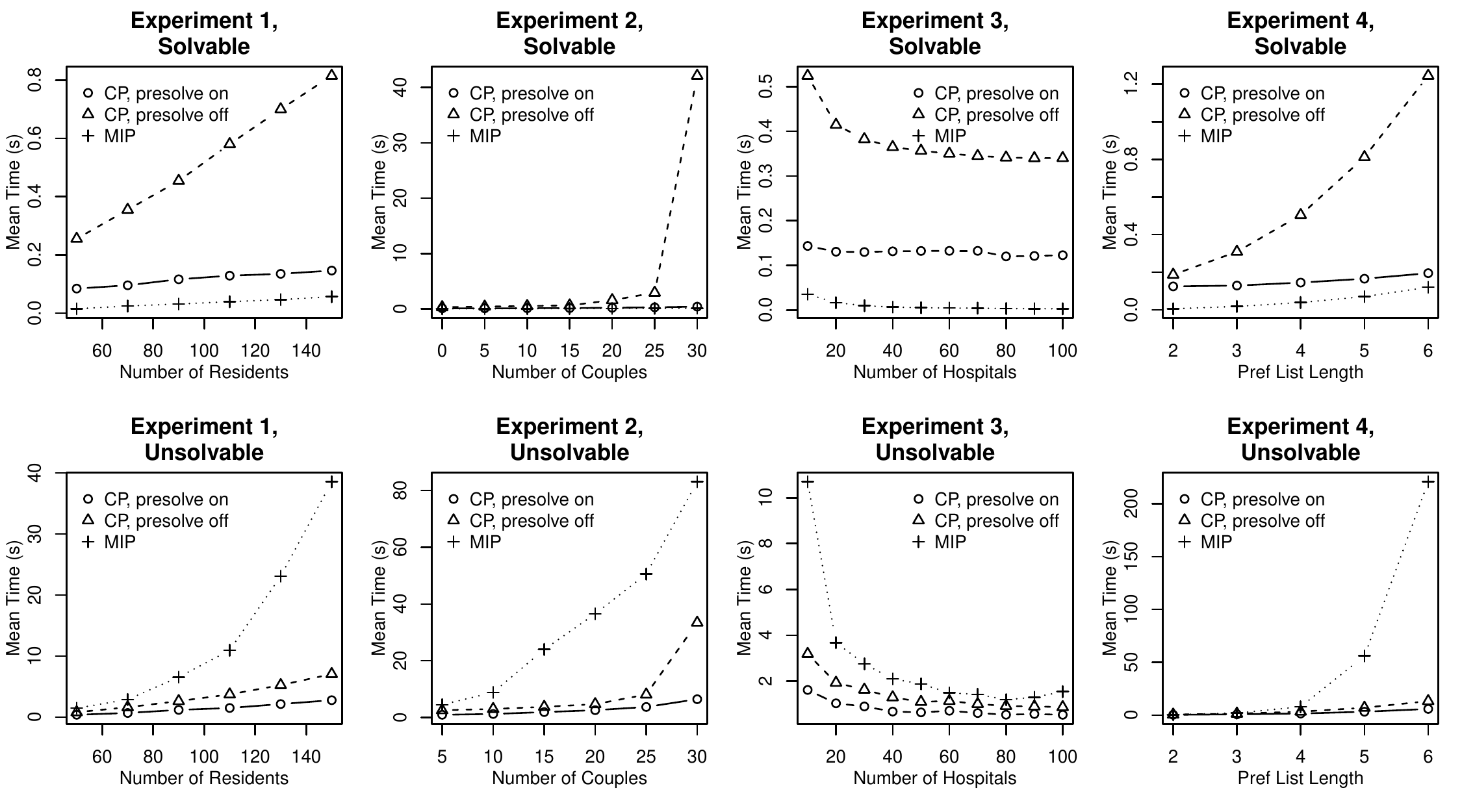}
\vspace{-5mm}
\caption{Comparison of run times using CP (with and without presolve) and MIP models}
\label{chart:cp_vs_mip}
\end{figure}

Figure~\ref{chart:cp_vs_mip} plots the mean run times for each of the four experiments for the IP model and for the CP models with and without presolving: each plot in the top row shows results for the solvable instances in one experiment, and each plot in the bottom row shows corresponding results for the unsolvable instances.  Table \ref{tab:summary} shows the actual mean and median runtimes for each model, taken over all 28,000 instances $\mathcal I$ across all four experiments, those instances from $\mathcal I$ that were solvable and those from $\mathcal I$ that were unsolvable.

\begin{table}
\begin{center}
\begin{tabular}{|r|c|c|c|c|c|c|}
\hline
Instance type & \multicolumn{3}{|c|}{Mean} & \multicolumn{3}{|c|}{Median}\\
\hline
& IP model & \multicolumn{2}{|c|}{CP model} & IP model & \multicolumn{2}{|c|}{CP model}\\
\cline{3-4} \cline{6-7}
& & No presolve & Presolve & & No presolve & Presolve\\
\cline{2-7}
All	& 2.568 & 2.237 & 0.315 & 0.031 & 0.430 & 0.129\\
Solvable & 0.034 & 1.839 & 0.143 & 0.016 & 0.402 & 0.127\\
Unsolvable & 30.781 & 6.669 & 1.276 & 8.948 & 3.240 & 1.395\\
\hline
\end{tabular}
\end{center}
\caption{Summary of mean and median runtimes over all experiments (all timings are in seconds)}
\label{tab:summary}
\end{table}

The CP model without presolve generally performs unfavourably for solvable instances.  Here, the IP model is faster than the CP model with presolve; this is likely to be due to the fact that for such instances, the earlier IP model for {\sc hrc} \cite{BMMcB14} is used instead of the more complex IP formulation for {\sc min bp hrc}.  For unsolvable instances, the CP model (with or without presolve) is faster than the IP model.  This is likely to be due to the fact that the CP model for {\sc min bp hrc} is more compact than its IP counterpart, involving fewer variables and constraints.  Comparing total run time summed across all 28,000 instances, the CP model was 1.15 times faster than the CP model without presolve, and the CP model with presolve was 8.14 times faster than the IP model.

When solving the CP model, the distribution of runtimes for the case without presolve had a very long right tail; 14 of the 28,000 instances accounted for over half of the total run time.  The longest-running instance took 17,617 seconds, and surprisingly this was a solvable instance (generated for Experiment 2).  For this reason, Table \ref{tab:summary} shows median run times as well as mean run times; from this we can see that the median runtime for the IP model is lower than that for the CP models for all instances and for solvable instances.  However for unsolvable instances, the median runtime for CP without presolve is 2.762 times faster than the median runtime for IP, and this factor increases to 6.414 for CP with presolve.

\section{Concluding remarks}
In this paper we have presented complexity and approximability results for {\sc min bp hrc}, showing that the problem is NP-hard and very difficult to approximate even in highly restricted cases.  We have then presented IP and CP models, together with empirical analyses of both models applied to randomly-generated {\sc hrc} instances.  Our main finding is that most-stable matchings admit a very small number of blocking pairs (in most cases at most 1, but never more than 2) on the instances we generated.  We also showed that on average the CP model is faster than the IP model, with the performance of the CP model being enhanced if presolving was carried out.  As far as future work is concerned, it would be interesting to determine the effect of presolving on the IP model, and more generally, to investigate further methods to enable the models to be scaled up to larger instances, such as column generation in the case of the IP model, and variable / value ordering heuristics in the case of the CP model.

\section*{Acknowledgements}
We would like to thank anonymous reviewers of an earlier version of this paper for their valuable comments, including the suggestion of references \cite{Sho00,Rob01,YGS03,Sol14}, which have helped to improve the presentation of this paper.
%\bibliographystyle{plain}
%\bibliography{matching}

%\input{Appendix}
\appendix

\chapter{Appendix}
%\chapter{Integer programming models for {\sc hrc} variants}
\section{Comparison of stability definitions}
\label{sec:stabdefcomp}
In this section we compare our stability definition for {\sc hrc} given by Definition \ref{stability:MM} with the definition adopted by Drummond et al.\ \cite{DPB15}.  Suppose that $I$ is an instance of {\sc hrc} and $M$ is a matching in $I$.  Let $R'\subseteq R$.  For a given hospital $h_j$, Drummond et al.\ defined $Ch_j(R')$ to be the set of residents that $h_j$ would select from $R'$.  That is, $Ch_j(R')$ is the maximal subset of $R'$ such that, for all $r_i\in Ch_j(R')$, $h_j$ finds $r_i$ acceptable, $h_j$ prefers $r_i$ to any $r_k\in R'\backslash Ch_j(R')$ and $|Ch(R')|\leq c_j$.  Then Drummond et al.\ defined the predicate {\sf willAccept}$(h_j,R',M)$ to be true if and only if $R'\subseteq Ch_j(M(h_j)\cup R')$.

Now suppose that $(r_i,r_j)$ is a couple in $I$ who prefer the hospital pair $(h_k,h_k)$ to their assigned hospital pair $(M(r_i),M(r_j))$, where $h_k$ is full in $M$. According to Condition 3(b) of the stability definition of Drummond et al.\ \cite{DPB15}, $h_k$ will participate in a blocking pair with $(r_i,r_j)$ if and only if {\sf willAccept}$(h_k,\{r_i,r_j\},M)$.  According to Condition 3(d) of Definition \ref{stability:MM}, $h_k$ will participate in a blocking pair with $(r_i,r_j)$ if and only if $h_k$ prefers $r_i$ to some $r_s\in M(h_k)$, and $h_k$ prefers $r_j$ to some $r_t\in M(h_k)\backslash \{r_s\}$.  Our Condition 3(d) is thus weaker than Condition 3(b) of Drummond et al.\ \cite{DPB15}, meaning that our stability definition is stricter.

To illustrate the difference, consider the {\sc hrc} instance $I$ shown in Figure \ref{fig:example}.  Here $h_1$ has capacity 2, whilst each of $h_2$ and $h_3$ has capacity 1.
\begin{figure}[h]
\begin{center}
\begin{tabular}[t]{rllccrrll}
\multicolumn{3}{l}{~~~~~~~~ Residents} &&& ~~~~~~~~~~~~~~ & \multicolumn{3}{l}{~~~~~~~~~~ Hospitals ~~~~~~~~~~~~} \\
\hline
$(r_1,r_2)$	&:  & $(h_1,h_1)$ ~ $(h_2,h_3)$   &&&&  $h_1$ &: $r_1 ~~ r_3 ~~ r_2 ~~ r_4$\\
$r_3$	    &:  & $h_1$                       &&&&  $h_2$ &: $r_1$\\
$r_4$	    &:  & $h_1$                       &&&&  $h_3$ &: $r_1$
\end{tabular}
\end{center}
\caption{An instance of {\sc hrc}.}
\label{fig:example}
\end{figure}

Let $M$ be the matching $\{(r_1,h_2),(r_2,h_3),(r_3,h_1),(r_4,h_1)\}$.  Then $(r_1,r_2)$ forms a blocking pair of $M$ with the hospital pair $(h_1,h_1)$ according to the stability definition given in Definition \ref{stability:MM}, but this does not happen with respect to the stability definition of Drummond et al.\ \cite{DPB15}.  In the latter case, $Ch_j(M(h_1)\cup \{r_1,r_2\})=Ch_j(\{r_1,r_2,r_3,r_4\})=\{r_1,r_3\}\not\supseteq \{r_1,r_2\}$.  We would argue that $(r_1,r_2)$ should form a blocking pair with $(h_1,h_1)$, because $h_1$ unequivocally improves by rejecting $\{r_3,r_4\}$ and taking on $\{r_1,r_2\}$ instead.

\section{Inapproximability result for $(\infty ,1, \infty)$-{\sc min bp hrc}}
\label{sec:inapprox}
We now establish that the problem of deciding whether an instance of \small$(\infty ,1, \infty)$\normalsize {\sc -hrc} admits a stable matching is NP-complete.
\begin{theorem1}
\label{213hrcfromvertexcover}
Given an instance of \small$(\infty ,1, \infty)$\normalsize {\sc -hrc}, the problem of deciding whether there exists a stable matching is NP-complete. 	The result holds even if each hospital has capacity 1.
\end{theorem1}

\begin{proof}
The proof of this result uses a reduction from a restricted version of the vertex cover problem.  More specifically, let {\sc vc3} denote the problem of deciding, given a cubic graph $G$ and an integer $K$, whether $G$ contains a vertex cover of size at most $K$.  This problem is NP-complete \cite{GJS76, MS77}.

The problem of deciding whether there exists a stable matching in an instance of \small$(\infty ,1, \infty)$\normalsize {\sc -hrc} is clearly in NP, as a given assignment may be verified to be a stable matching in polynomial time. To show NP-hardness, let $\langle G,K \rangle$ be an instance of {\sc vc3}, where $G = ( V,E)$, $E = \{e_1 , \ldots, e_m\}$ and $V = \{v_1 , \ldots , v_n\}$. For each $i$ $(1\leq i\leq n)$, suppose that $v_i$ is incident to edges $e_{j_1}$, $e_{j_2}$ and $e_{j_3}$ in $G$, where without loss of generality $j_1 < j_2 < j_3$. Define $e_{i,s} = e_{j_s}$ $(1\leq s\leq 3)$. Similarly, for each $j$ $(1\leq j\leq m)$, suppose that $e_j = \{v_{i_1} , v_{i_2}\}$, where without loss of generality $i_1 < i_2 $. Define $v_{j,r} = v_{i_r}$ $(1 \leq r\leq 2)$.

\begin{figure}
\[
\begin{array}{rll}

\multicolumn{3}{c}{\mbox{Residents' Preferences}} \\
\hline 
\\
( r_j^1 , r_j^2) : & ( h_j^1 , h_j^2) & (1\leq j\leq m) \vspace{1mm}\\
( r_j^3 , r_j^4) : & ( h_j^1 , h_j^2) & (1\leq j\leq m) \vspace{1mm}\\
\\
(f_t^1, f_t^2) : & (g_t^1, g_t^2) & (1\leq t\leq K) \vspace{1mm} \\ 
(f_t^3, f_t^4) : & (g_t^2, g_t^3) & (1\leq t\leq K) \vspace{1mm} \\ 
(f_t^5, f_t^6) : & (g_t^3, g_t^1) & (1\leq t\leq K) \vspace{1mm} \\ 
\\
(y_t^1, y_t^2) : & (z_t^1, z_t^2) & (1\leq t\leq n-K) \vspace{1mm} \\ 
(y_t^3, y_t^4) : & (z_t^2, z_t^3) & (1\leq t\leq n-K) \vspace{1mm} \\ 
(y_t^5, y_t^6) : & (z_t^3, z_t^1) & (1\leq t\leq n-K) \vspace{1mm} \\ 
\\
a_t : & p_t ~~ g_t & (1\leq t\leq K) \vspace{1mm}\\
b_t : & q_t ~~ z_t & (1\leq t\leq n-K) \vspace{1mm}\\
\\
x_i : & p_1 ~~ p_2 ~~ \ldots ~~ p_K ~~ h^1(x_i) ~~ h^2(x_i) ~~ h^3(x_i) ~~ q_{1} ~~ q_{2} ~ \ldots ~~ q_{n-K} & (1\leq i\leq n) \vspace{1mm}\\
\\
\multicolumn{3}{c}{\mbox{Hospitals' Preferences}} \\
\hline
\\
g_t^1 : & a_t ~~ f_t^1 ~~ f_t^6 & (1\leq t\leq K)  \vspace{1mm}\\
g_t^2 : & f_t^3 ~~ f_t^2 & (1\leq t\leq K)  \vspace{1mm}\\
g_t^3 : & f_t^5 ~~ f_t^4 & (1\leq t\leq K)  \vspace{1mm}\\
\\
h_j^1 : & r_j^1 ~~ x(h_j^1) ~~ r_j^3 & (1\leq j\leq m) \vspace{1mm}\\
h_j^2 : & r_j^4 ~~ x(h_j^2) ~~ r_j^2 & (1\leq j\leq m) \vspace{1mm}\\
\\
%p_t : & x_{1} ~~ x_{2} ~~ \ldots ~~ x_{n} & (1\leq t\leq K) \vspace{1mm}\\
%q_t : & x_{1} ~~ x_{2} ~~ \ldots ~~ x_{n} & (K+1\leq t\leq n) \vspace{1mm}\\
p_t : & x_{1} ~~ x_{2} ~~ \ldots ~~ x_{n} ~~ a_t & (1\leq t\leq K) \vspace{1mm}\\
q_t : & x_{1} ~~ x_{2} ~~ \ldots ~~ x_{n} ~~ b_t & (1\leq t\leq n-K) \vspace{1mm}\\
\\
z_t^1 : & b_t ~~ y_t^1 ~~ y_t^6 & (1\leq t\leq n-K) \vspace{1mm}\\
z_t^2 : & y_t^3 ~~ y_t^2 & (1\leq t\leq n-K) \vspace{1mm}\\
z_t^3 : & y_t^5 ~~ y_t^4 & (1\leq t\leq n-K) \vspace{1mm}\\
\\
\end{array}
\]
\caption{Preference lists in $I$, the constructed instance of \small $(\infty,1,\infty)$\normalsize {\sc -hrc}.}
\label{preflistsinf1infhrc}
\end{figure}

We form an instance $I$ of \small $(\infty ,1, \infty)$\normalsize {\sc -hrc} as follows. The set of residents in $I$ is $A \cup B \cup F\cup R\cup X \cup Y$ where  $A=\{a_t : 1\leq t\leq K\}$, $B=\{b_t : 1\leq t\leq n-K\}$, $F=\bigcup_{t=1}^{K} F_t$, where $F_t=\{f_t^s : 1\leq s\leq 6\} $, $R=\bigcup_{j=1}^{m} R_j$, where $R_j=\{r_j^s : 1\leq s\leq 4\}$, $X=\{x_i : 1\leq i\leq n\}$ and $Y=\bigcup_{t=1}^{n-K} Y_t$, where $Y_t=\{y_t^s : 1\leq s\leq 6\}$.
 
The set of hospitals in $I$ is $G\cup H\cup P\cup Q\cup Z$, where $G=\bigcup_{t=1}^{K} G_t$, where $G_t=\{g_t^r : 1\leq r\leq 3\}$ $ (1\leq t\leq K)$, $H=\bigcup_{j=1}^{m} H_j$, $H_j=\{h_j^s : 1\leq s\leq 2\}$ , $P=\{p_t : 1\leq t\leq K\}$, $Q=\{q_t : 1\leq t\leq n-K\}$ and $Z=\bigcup_{t=1}^{n-K} Z_t$, where $Z_t=\{z_t^r : 1\leq r\leq 3\}$ and each hospital has capacity 1. The preference lists of the resident couples, single residents and hospitals in $I$ are shown in Figure \ref{preflistsinf1infhrc}. 

In the preference list of a resident $x_i$ $(1\leq i\leq n)$ the symbol $h^s(x_i)$ $(1\leq s\leq 3)$ denotes the hospital $h_j^r$ $(1\leq r\leq 2)$ such that $e_j = e_{i,s}$ and $v_i = v_{j,r}$. Similarly, in the preference list of a hospital $h_j^r$ $(1\leq j\leq m, 1\leq r\leq 2)$ the symbol $x(h_j^r)$ denotes the resident $x_i$ such that $v_i = v_{j,r}$.

We claim that $G$ contains a vertex cover of size at most $K$ if and only if $I$ admits a stable matching. Let $C$ be a vertex cover in $G$ such that $ |C|  \leq K$. Without loss of generality we may assume that $|C| = K$ for if otherwise a sufficient number of vertices can be added to $C$ without violating the vertex cover condition.

We show how to define a matching $M$ in $I$ as follows. Let $C = \{v_{r_1} , v_{r_2}, \ldots , v_{r_K}\}$ where without loss of generality $r_1 < r_2 < \ldots < r_K$. Further let $V \setminus C = \{v_{s_1} , v_{s_2}, \ldots , v_{s_{n-K}}\}$ where without loss of generality $s_1 < s_2 < \ldots < s_{n-k}$. For each vertex $v_{r_i} \in C$ add the pairs $\{(x_{r_i} , p_i), (a_i , g_i), (f_i^3 , g_i^2), (f_i^4 , g_i^3)\}$ for $1\leq i\leq K$ to $M$. For each vertex $v_{s_i} \in V \setminus C$ add $\{(x_{s_i} , q_i), (b_i , z_i), (y_i^3 , z_i^2), (y_i^4 , z_i^3)\}$ for $1\leq i\leq n-K$ to $M$. 

For each edge $e_j \in E$ at least one of $v_{j,1}$ or $ v_{j,2}$ must be in $C$. If $v_{j,1} \in C$ add the pairs $\{(r_j^3 , h_j^1) , (r_j^4 , h_j^2)\}$ to $M$. Otherwise $v_{j,2} \in C$ so  add the pairs $\{(r_j^1 , h_j^1) , (r_j^2 , h_j^2)\}$ to $M$.

We now show that $M$ is a stable matching in $I$.  Firstly, we show that no hospital $h_j^r \in H$ $(1\leq j\leq m, 1\leq r\leq 2)$ can form part of a blocking pair of $M$. Assume a hospital $h_j^r \in H$ is part of a blocking pair of $M$ for some $j ~~ (1\leq j\leq m)$ and $r ~~ (1\leq r\leq 2)$. Now, since $C$ is a vertex cover in $G$, an arbitrary edge $e_j \in E$ must covered by either $v_{j,1}$ or $v_{j,2}$ or both. Assume firstly that $v_{j,1} \in C$. Then by construction $(x_{r_t}, p_{t}) \in M$ and $\{(r_j^3 , h_j^1) , (r_j^4 , h_j^2)\} \subseteq M$ where $v_{j,1} = v_{r_t}$. Assume $(  x(h_j^1) , h_j^1)$ blocks $M$ for some $j ~~ (1\leq j\leq m)$. Since $v_{j,1} \in C$ and thus $M(x(h_j^1)) \in P$, $x(h_j^1)$ prefers $M(x(h_j^1))$ to $h_j^1$, a contradiction. Now assume that $(  (r_j^1 , r_j^2), ( h_j^1 , h_j^2))$ blocks $M$. However, $h_j^2$ prefers $M(h_j^2) = r_j^4$ to $r_j^2$, a contradiction. 

Now assume $v_{j,1} \notin C$. Then $v_{j,2} \in C$ and by construction $(x_{r_{t^{\prime }}}, p_{t^{\prime }}) \in M$ and $\{(r_j^1 , h_j^1),$ $ (r_j^2 , h_j^2)\} \subset M$ where $v_{j,2} = v_{r_{t^{\prime }}}$. Assume $(  x(h_j^2) , h_j^2)$ blocks $M$ for some $j ~~ (1\leq j\leq m)$. Since $v_{j,2} \in C$ and thus $M(x(h_j^2)) \in P$, $x(h_j^2)$ prefers $M(x(h_j^2))$ to $h_j^2$, a contradiction. Now assume $( (r_j^3 , r_j^4), ( h_j^1 , h_j^2))$ blocks $M$. However, $h_j^1$ prefers $M(h_j^1) = r_j^1$ to $r_j^3$, a contradiction. Thus, no $h_j^r \in H$ $(1\leq j\leq m, 1\leq r\leq 2)$ can form part of a blocking pair of $M$.

We now show that no hospital in $P \bigcup Q$ can be involved in a blocking pair of $M$. By construction $M(p_t) \in X$ for all $t ~~ (1\leq t\leq K)$. Assume some pair $( x_{k_1} , p_{l_1})$ blocks $M$. Let $M(x_{k_1}) = p_{l_2}$ and $M( p_{l_1}) = x_{k_2}$.  Since $( x_{k_1} , p_{l_1})$ blocks $M$ then $l_1 < 1_2$ and $k_1 < k_1$ in contradiction to the construction of $M$. A similar argument shows that no hospital in $Q$ may be involved in a blocking pair of $M$ and thus we have that no hospital in $P\bigcup Q$ may be involved in a blocking pair of $M$.

We now show that no hospital in $G \bigcup Z$ can be involved in a blocking pair of $M$. Firstly, assume a hospital $g_t^s \in H$ is part of a blocking pair of $M$ for some $t ~~ (1\leq t\leq K)$ and $s ~~ (1\leq s\leq 3)$. Clearly, since $g_t^1$ and $g_t^2$ are both assigned their first preference they cannot form part of a blocking pair for $M$. Hospital $g_t^3$ prefers $f_t^5$ to $M(g_t^3) = f_t^4$. However, $f_t^5$ is a member of the couple $( f_t^5 , f_t^6)$ that expresses a joint preference for the pair $( g_t^3 , g_t^1)$ and $g_t^1$ prefers $M (g_t^1) = a_t$ to $f_t^6$, a contradiction. Thus, no hospital $g_t^s \in H$ $(1\leq t\leq K, 1\leq s\leq 3)$ can form part of a blocking pair of $M$. A similar argument may be used to show that no $z_t^s \in H$ $(1\leq t\leq n-K, 1\leq s\leq 3)$ can form part of a blocking pair of $M$ and thus we have that no hospital in $G \bigcup Z$ can be involved in a blocking pair of $M$.

We now have that no hospital in $I$ may be part of a blocking pair of $M$ and thus $M$ must be stable.

\medskip

Conversely, let $M$ be a stable matching in $I$. 
We first show that the stability of $M$ implies that $M(x_i) \in P \bigcup Q$ for all $i$ $(1\leq i\leq n)$. Observe that if $(a_t , g_t^1) \notin M$ for $t$ $(1\leq t\leq K)$ then no stable assignment is possible amongst the agents in $F_t \bigcup G_t$ as shown in Lemma \ref{lemma_212HRC_sub-instance_stableifeven}. However, if $\{(a_t , g_t^1),$ $(f_t^3 , g_t^2),$ $(f_t^4 , g_t^3)\} \subseteq M$ then no blocking pair exists in $F_t \bigcup G_t$. It follows that if  $(a_t , g_t^1) \in M$ then $(a_t , p_t)$ blocks $M$ unless $M(p_t) \in X$. A similar argument shows that $M(q_t) \in X$ for all $t$ $(1\leq t\leq n-K)$. Now, since $|X| = n$ and $|P \bigcup Q | = n$, clearly all $x \in X$ must be partnered with a member of $P \bigcup Q$ and moreover, $M(x_i) \notin H$ in any stable matching in $I$.

Next we show that the stability of $M$ implies that $h_j^1$ and $h_j^2$ are fully subscribed in $M$ for all $j$ $(1\leq j\leq m)$. Let $j$ $(1\leq j\leq m)$ be given. Assume that both $h_j^1$ and $h_j^2$ are undersubscribed in $M$. Since $M(x(h_j^r)) \neq h_j^r$ for all $j, r$ $(1\leq j\leq m, 1\leq r\leq 2)$, $( (r_j^1 , r_j^2), (h_j^1 , h_j^2))$ blocks $M$, a contradiction. Thus either $\{(r_j^1 , h_j^1), ( r_j^2, h_j^2))\} \subseteq M$ or $\{(r_j^3 , h_j^1), ( r_j^4, h_j^2))\} \subseteq M$ in any stable matching in $I$. If $\{(r_j^1 , h_j^1), ( r_j^2, h_j^2))\} \subseteq M$ then $( (r_j^3 , r_j^4), (h_j^1 , h_j^2))$ does not block $M$. Similarly, if $\{(r_j^3 , h_j^1), ( r_j^4, h_j^2))\} \subseteq M$ then $( (r_j^1 , r_j^2), (h_j^1 , h_j^2))$ does not block $M$. Thus we have that $h_j^1$ and $h_j^2$ are fully subscribed in $M$ for all $j$ $(1\leq j\leq m)$. Moreover, we have that all hospitals must be fully subscribed in any stable matching $M$ in $I$.

Define a set of vertices $C$ in $G$ as follows. For each $i~ (1\leq i\leq n)$ if $M(x_i) \in P$, add $v_i$ to $C$. Since $M$ is a stable matching and $|P| = K$, this process selects exactly $K$ of the $n$ vertices in $V$ and thus $|C| = K$. We now show that $C$ represents a vertex cover in $G$. Consider an arbitrary edge $e_j \in E$. Assume that both $v_{j,1} \notin C$ and $v_{j,2} \notin C$ and hence that $C$ is not a vertex cover in $G$. Then $M(x_{j,1}) \in Q$ and $M(x_{j,2}) \in Q$. As $M$ is stable and thus hospital complete, either $\{(r_j^1 , h_j^1) , (r_j^2 , h_j^2)\} \subset M$ or $\{(r_j^3 , h_j^1) , (r_j^4 , h_j^2)\} \subset M$. If $\{(r_j^1 , h_j^1) , (r_j^2 , h_j^2)\} \subset M$ then $( x_{j,2} , h_j^2)$ blocks $M$, a contradiction.  If $\{(r_j^3 , h_j^1) , (r_j^4 , h_j^2)\} \subset M$ then $( x_{j,1} , h_j^1)$ blocks $M$, a contradiction. Hence $C$ represents a vertex cover in $G$ of size $K$ and the theorem is proven. \qed
\end{proof}
\begin{corollary1}
\label{inapprox}
\small$(\infty ,1, \infty)$\normalsize {\sc -min bp hrc} is NP-hard and not approximable within a factor of $n_1^{1-\varepsilon}$, for any $\varepsilon>0$, unless P=NP, where $n_1$ is the number of residents in a given instance.  The result holds even if each hospital has capacity 1.
\end{corollary1}
\begin{proof}
The proof of this result is analogous to the proof of Theorem 2 in \cite{McBM13}, which establishes the same result for \small$(0,2,2)$\normalsize {\sc -min bp hrc}, using the NP-complete\-ness of the problem of deciding whether a stable matching exists in a given instance of \small$(0,2,2)$\normalsize {\sc -hrc} as a starting point.  The restrictions on the preference list lengths are not used in the gap-introducing reduction that proves the inapproximability result, hence the same reduction can be used to demonstrate the inapproximability of \small$(\infty ,1, \infty)$\normalsize {\sc -min bp hrc}. \qed
\end{proof}
%\begin{figure}
%\[
%\begin{array}{rll}
%
%\multicolumn{3}{c}{Residents} \\
%\hline 
%\\
%( x_j^1 , r_j^1) : & ( h(x_j^1) , y_j^1) & (1\leq j\leq m) \vspace{1mm}\\
%( x_j^2 , r_j^2) : & ( h(x_j^2) , y_j^2) & (1\leq j\leq m) \vspace{1mm}\\
%( r_j^3 , r_j^4) : & ( y_j^1 , y_j^2) & (1\leq j\leq m) \vspace{1mm}\\
%( r_j^5 , r_j^6) : & ( y_j^2 , y_j^1) & (1\leq j\leq m) \vspace{1mm}\\
%
%
%\\
%(p_i^1, p_i^2) : & (h_i^1, h_i^2) & (1\leq i\leq n) \vspace{1mm} \\ 
%(p_i^3, p_i^4) : & (h_i^2, h_i^3) & (1\leq i\leq n) \vspace{1mm} \\ 
%(p_i^5, p_i^6) : & (h_i^3, h_i^1) & (1\leq i\leq n) \vspace{1mm} \\ 
%\\
%\multicolumn{3}{c}{Hospitals} \\
%\hline
%\\
%h_i^1 : & x(h_i^1) ~ p_i^1 ~ p_i^6 & (1\leq i\leq n) \vspace{1mm}\\
%h_i^2 : & x(h_i^2) ~ p_i^3 ~ p_i^2 & (1\leq i\leq n) \vspace{1mm}\\
%h_i^3 : & x(h_i^3) ~ p_i^5 ~ p_i^4 & (1\leq i\leq n) \vspace{1mm}\\
%\\
%y_j^1 : & r_j^3 ~ r_j^1 ~ r_j^4 & (1\leq i\leq n) \vspace{1mm}\\
%y_j^2 : & r_j^5 ~ r_j^2 ~ r_j^6  & (1\leq i\leq n) \vspace{1mm}\\
%\\
%\end{array}
%\]
%\caption{Preference lists in $I$, the constructed instance of \small $(2,1,3)$\normalsize {\sc -hrc}.}
%\label{preflists213hrc}
%\end{figure}

\section{Efficiently solvable variants of {\sc min bp hrc}}
\label{sec:algs}
\subsection{Fixed assignments in {\sc hrc}}
\label{section:fixed_assignments}

In an instance $I$ of {\sc hrc} some agents may rank one another highly in their preference lists, leading to the outcome that they must be assigned to one another in any stable matching in $I$.  We describe these agents as \emph{fixed assignments}, using the following lemma to define this concept formally and to show that fixed assignments must belong to any stable matching in $I$.

\begin{lemma1}
\label{lemma:fixedAssignments}
Let $I$ be an arbitrary instance of {\sc hrc}. 
\begin{itemize}
\item[(i)] If a single resident $r_i$ has a hospital $h_j$ in first place on his preference list and $r_i$ is within the first $c_j$ places on $h_j$'s preference list then $(r_i, h_j)$ must belong to any stable matching in $I$.
\item[(ii)] If a couple $(r_{i}, r_j)$ has a hospital pair $(h_{p}, h_q)$ in first place on its joint preference list and $r_i$ is within the first $c_p$ places on $h_p$'s preference list and also $r_j$ is within the first $c_q$ places on $h_q$'s preference list then $(r_i, r_j)$ must be jointly assigned to $(h_p, h_q)$ in any stable matching in $I$. 
\end{itemize}
Any pair consisting of a single resident and a single hospital satisfying Case (i), or consisting of a couple and hospital pair satisfying Case (ii), is called a \emph{fixed assignment} in $I$.
\end{lemma1}
\begin{proof}
The proof of the Lemma follows immediately from the fact that any matching $M$ in which $r_i$ is not assigned to $h_j$ will be blocked by $(r_i, h_j)$, and similarly, any matching $M$ in which $(r_i, r_j)$ and $(h_p, h_q)$ are not assigned to each other will be blocked by $(r_i, r_j)$ with $(h_p, h_q)$. \qed
\end{proof}

Suppose that a matching $M$ is constructed solely by matching agents who are involved in fixed assignments.  As a consequence, suppose some pair $(r_i,h_j)$ is added to $M$.  Clearly no other hospital may be assigned in $M$ to $r_i$, and hence $r_i$ can be deleted from the preference list of each other hospital in which he appears. Moreover, in the event that $h_j$ becomes fully subscribed by accepting $r_i$ as an assignee, $h_j$ can be deleted from the preference list of each resident other than $r_i$ in which it appears.  We say that we \emph{satisfy} a fixed assignment if we match together in $M$ the agents involved, and then carry out the corresponding deletions as described above.  Note that making these deletions may expose another fixed assignment in the resulting reduced instance of {\sc hrc}, which can then also be satisfied in $M$.  If we continue satisfying fixed assignments until no more fixed assignments are exposed then we say all fixed assignments have been \emph{iteratively satisfied} in $M$.  This idea is used in the proof of the following proposition, to show that a stable solution can be found in an instance of {\small$(\infty , \infty , 1)$\normalsize {\sc -hrc} in polynomial time. 

\begin{proposition1}
\label{proposition:221hrc}
An instance $I$ of \small$(\infty , \infty , 1)$\normalsize {\sc -hrc} admits exactly one stable matching, which can be found in polynomial time.
\end{proposition1}

\begin{proof}
Consider an arbitrary single resident $r_i$ in $I$. Let the hospital in first place on resident $r$'s preference list be $h_j$. Since $r_i$ must be in first place in $h_j$'s preference list (as it is the only preference expressed by $h_j$), the pair $(r_i,h_j)$ represents a fixed assignment in $I$. Thus, any single resident in $I$ must be part of exactly one fixed assignment in $I$ and this may be satisfied by assigning each single resident to the hospital in first place on his preference list.

Now, consider an arbitrary couple $(r_i, r_j)$ in $I$. Let the hospital pair $(h_p , h_q)$ be in first place on couple $(r_i, r_j)$'s joint preference list. Clearly, since $r_i$ (respectively $r_j$) is in first place on $h_p$'s (respectively $h_q$'s) preference list, $(r_i, r_j)$ with $(h_p , h_q)$ represents a fixed assignment in $I$. Thus, any resident couple in $I$ must be part of exactly one fixed assignment in $I$ and this may be satisfied by assigning each couple to the hospital pair in first place on their joint preference list.

Hence, the fixed assignments involving both the single residents and the couples in $I$ may be satisfied iteratively in time linear in the number of residents in $I$, leading to a matching $M$ that is clearly stable in $I$, and which is the only stable matching in $I$. \qed
\end{proof}

\subsection{\small$(2,1,2)$\normalsize{\sc -min bp hrc} is efficiently solvable}
\label{section:212hrc_efficiently_solvable}

%Proposition \ref{proposition:221hrc} established that the fixed assignments in an arbitrary instance of \hyperref[section:introduction_hrc]{{\sc hrc}} may be satisfied in polynomial time. 
Let $I$ be an instance of {\small$(2,1,2)$\normalsize {\sc -hrc}}, and assume that $M_0$ is a matching in $I$ in which all fixed assignments have been iteratively satisfied, and assume that the corresponding deletions have been made from the preference lists in $I$.  In Lemma \ref{lemma_212HRC_sub-instance} below, we use the absence of fixed assignments in $I$ to infer that $I$ must be constructed from the union of a finite number of disjoint discrete sub-instances of \small$(2,1,2)$\normalsize {\sc -hrc} and further that each disjoint sub-instance $I'$ of $I$  must be of the form shown in Figure \ref{preflists:the full instance}. Let $I'$ be one of these disjoint sub-instances of $I$. We prove in Lemma \ref{lemma_212HRC_sub-instance_stableifeven} that the number of couples involved in $I'$ determines whether $I'$ admits a stable matching: indeed, $I'$ admits a stable matching if and only if the number of couples in $I'$ is even.

\begin{lemma1}
\label{lemma_212HRC_sub-instance}
An arbitrary instance of {\small$(2,1,2)$\normalsize {\sc -hrc}} involving at least one couple and in which all fixed assignments have been iteratively satisfied must be constructed from sub-instances of the form shown in Figure \ref{preflists:the full instance} in which all of the hospitals have capacity 1.
\end{lemma1}

\begin{proof}

Let $I$ be an arbitrary instance of {\small$(2,1,2)$\normalsize {\sc -hrc}} in which all fixed assignments have been iteratively satisfied. Observe that if a couple expresses a preference for a hospital pair $(h_p , h_p)$ this would represent a fixed assignment, a contradiction. Thus, no couple may express such a preference in $I$. We now show how the absence of fixed assignments in $I$ allows us to infer the preference lists for all of the agents involved in $I$.

\begin{figure}[t]
\begin{center}
\begin{tabular}[t]{rllccrrll}

 \multicolumn{3}{l}{~~~~~~~~ Residents} &&& ~~~~~~~~ & \multicolumn{3}{l}{~~~~~~~~~~ Hospitals ~~~~~~~~~~~~} \\
\hline

\\

 $(r^1_{c_1}, r^2_{c_1})$ & : & $(h^0_{c_1}, h^1_{c_1})$ 	&&&& $h^0_{c_1}$ &: ~$r^1_{c_1}$ ~ $r^{n_N}_{s_N}$\\ 

\smallskip

 $r^1_{s_1}$  &:  & $h^2_{c_1}$ ~ $h^1_{c_1}$ 	&&&&  $h^1_{c_1}$ &: ~$r^1_{s_1}$ ~ $r^2_{c_1}$\\
 $r^2_{s_1}$  &:	& $h^3_{c_1}$ ~ $h^2_{c_1}$ 	&&&&  $h^2_{c_1}$ &: ~$r^2_{s_1}$ ~ $r^1_{s_1}$	\\
\smallskip
 & $\vdots$  &&&&& $\vdots$ \\
\smallskip
 $r^{n_1}_{s_1}$ &: & $h^{{n_1}+1}_{c_1}$ ~ $h^{n_1}_{c_1}$ 	&&&& $h^{n_1}_{c_1}$	&:	~$r^{n_1}_{s_1}$ ~ $r^{n_1 - 1}_{s_1}$	\\ 
\smallskip
 $(r^1_{c_2}, r^2_{c_2})$ & : & $(h^{{n_1}+1}_{c_1}, h^1_{c_2})$ &&&&  $h^{{n_1}+1}_{c_1}$	&:	~$r^{1}_{c_2}$ ~ $r^{n_1}_{s_1}$	\\
\smallskip
 $r^1_{s_2}$	&:  & $h^2_{c_2}$ ~ $h^1_{c_2}$ 	&&&&  $h^1_{c_2}$ &: ~$r^1_{s_2}$ ~ $r^{2}_{c_2}$	\\
 $r^2_{s_2}$	&:  & $h^3_{c_2}$ ~ $h^2_{c_2}$ 	&&&&  $h^2_{c_2}$ &: ~$r^2_{s_2}$ ~ $r^1_{s_2}$\\
 & $\vdots$  &&&&& $\vdots$ \\
\smallskip
 $r^{n_2}_{s_2}$ &: & $h^{{n_2}+1}_{c_2}$ ~ $h^{n_2}_{c_2}$ 	&&&& $h^{n_2}_{c_2}$	&:	~$r^{n_2}_{s_2}$ ~ $r^{n_2 - 1}_{c_2}$	\\ 
\smallskip
 $(r^1_{c_3}, r^2_{c_3})$ & : & $(h^{{n_2}+1}_{c_2}, h^1_{c_3})$ &&&& $h^{{n_2}+1}_{c_2}$ &: ~$r^1_{c_3}$ ~ $r^{n_2}_{s_2}$	 	\\
 \smallskip
 $r^1_{s_3}$	&:  & $h^2_{c_3}$ ~ $h^1_{c_3}$ 	&&&&  $h^1_{c_3}$ &: ~$r^1_{s_3}$ ~ $r^2_{c_3}$\\
 $r^2_{s_3}$	&:  & $h^3_{c_3}$ ~ $h^2_{c_3}$ 	&&&&  $h^2_{c_3}$ &: ~$r^2_{s_3}$ ~ $r^1_{s_3}$\\
 & $\vdots$  &&&&& $\vdots$ \\

 $r^{n_{N-1}}_{s_{N-1}}$ &: & $h^{{n_{N-1}}+1}_{c_{N-1}}$ ~ $h^{n_{N-1}}_{c_{N-1}}$	&&&& $h^{n_{N-1}}_{c_{N-1}}$	&:	~$r^{n_{N-1}}_{s_{N-1}}$ ~ $r^{n_{N-1} - 1}_{c_{N-1}}$	\\ 
\smallskip
 $(r^1_{c_N}, r^2_{c_N})$ & : & $(h^{{n_{N-1}}+1}_{c_{N-1}}, h^1_{c_N})$ &&&& $h^{{n_{N-1}}+1}_{c_{N-1}}$ &: ~$r^1_{c_N}$ ~ $r^{n_{N-1}}_{s_{N-1}}$	 	\\
 \smallskip
 $r^1_{s_N}$	&:  & $h^2_{c_N}$ ~ $h^1_{c_N}$ 	&&&&  $h^1_{c_N}$ &: ~$r^1_{s_N}$ ~ $r^2_{c_N}$\\
 $r^2_{s_N}$	&:  & $h^3_{c_N}$ ~ $h^2_{c_N}$ 	&&&&  $h^2_{c_N}$ &: ~$r^2_{s_N}$ ~ $r^1_{s_N}$\\
 & $\vdots$  &&&&& $\vdots$ \\
 $r^{n_N}_{s_N}$	&:  & $h^0_{c_1}$ ~ $h^{n_N}_{c_N}$ 	&&&&  $h^{n_N}_{c_N}$ &: ~$r^{n_N}_{s_N}$ ~ $r^{n_{N-1}}_{s_N}$\\

\end{tabular}
\end{center}
\caption{An instance of \small$(2,1,2)$\normalsize {\sc -hrc} containing an arbitrary number of couples and an arbitrary number of residents that has no unsatisfied fixed assignments.}
\label{preflists:the full instance}
\end{figure}

\begin{figure}[t]
\begin{center}
\begin{tabular}[t]{lclclll}
\multicolumn{7}{c}{Residents} \\
\hline
\\

\medskip
$(r^1_{c_1}, r^2_{c_1})$		&	:		&	~ $(h^0_{c_1}, h^1_{c_1})$ 	& ~~~~ ~~~~ & 		&		& 	\\

\medskip

$(r^1_{c_k}, r^2_{c_k})$		&	:		&	~ $(h^{n_{k-1}+1}_{c_{k-1}}, h^1_{c_k})$ 	& & $2\leq k\leq N-1$			&		& 	\\

\medskip

$(r^1_{c_N}, r^2_{c_N})$		&	:		&	~ $(h^{n_{N-1}+1}_{c_{N-1}}, h^1_{c_N})$ 	& & $n_N>0$			&		& 	\\

\medskip

$(r^1_{c_N}, r^2_{c_N})$		&	:		&	~ $(h^{n_{N-1}+1}_{c_{N-1}}, h^0_{c_1})$ 	& & $n_N=0$			&		& 	\\

\medskip

 ~~~~ $r^a_{s_k}$		&	:	& ~ $h^{a+1}_{c_k}$ ~ $h^a_{c_k}$ 	&	 	& $1\leq k\leq N , 1 \leq a \leq n_k , n_k >0$	&		& 	\\

\\

\multicolumn{7}{c}{Hospitals} \\
\hline
\\
\medskip
 ~~~~ $h^0_{c_1}$ 	&	:	&	~ $r^1_{c_1}$ ~ $r^{n_N}_{s_N}$	& & if $n_N >0$	& &	\\
 
 \medskip
 
  ~~~~ $h^0_{c_1}$ 	&	:	&	~ $r^1_{c_1}$ ~ $r^{2}_{c_N}$	& & if $n_N = 0$	& &	\\
 
% \medskip
% 
% ~~~~ $h^1_{c_1}$ 	&	:	&	~ $r^1_{s_1}$ ~ $r^{2}_{c_1}$	& &	 if $n_1 > 0$& &	\\
% 
% \medskip
% 
% ~~~~ $h^1_{c_1}$ 	&	:	&	~ $r^1_{c_2}$ ~ $r^{2}_{c_1}$	& &  if $n_1 = 0$	& &	\\
 
 \medskip
 
 ~~~~ $h^1_{c_k}$ 	&	:	&	~ $r^1_{s_k}$ ~ $r^{2}_{c_{k}}$	& & $1\leq k\leq N$ , if  $n_k > 0$	& &	\\

\medskip

 ~~~~ $h^1_{c_k}$ 	&	:	&	~ $r^1_{c_{k+1}}$ ~ $r^{2}_{c_{k}}$	& & $1\leq k\leq N$, if $n_k = 0$	& &	\\

 \medskip
 ~~~~ $h^a_{c_k}$ 	&	:	&	~ $r^a_{s_k}$ ~ $r^{a-1}_{s_k}$	& & $1\leq k\leq N$, $2\leq a\leq n_k$, if $n_k >0$	& &	\\

 \medskip
 ~~~~ $h^{n_k+1}_{c_k}$ 	&	:	&	~ $r_{c_{k+1}}^1 $ ~ $r_{s_k}^{n_k}$	& & $1\leq k\leq N-1$, if $n_k >0$	& &	
\end{tabular}
\end{center}
\caption{An exactly equivalent description of the instance shown in Figure \ref{preflists:the full instance}}
\label{preflists:the formal instance}
\end{figure}

Let $(r^1_{c_1}, r^2_{c_1})$ be a couple in $I$ and further let $(h^0_{c_1}, h^1_{c_1})$ be the hospital pair for which $(r^1_{c_1}, r^2_{c_1})$ expresses a preference. Since all fixed assignments have been iteratively satisfied by construction, it cannot be the case that \emph{both}:
\begin{itemize}
\item[(i)] $h^0_{c_1}$ has capacity two or has $r^1_{c_1}$ in first place in its preference list \emph{and}
\item[(ii)] $h^1_{c_1}$ has capacity two or has $r^2_{c_1}$ in first place in its preference list.
\end{itemize}

Without loss of generality, assume that $h^1_{c_1}$ has capacity one and does not have $r^2_{c_1}$ in first place in its preference list. Hence there exists some other resident $r_x$ who is preferred by $h^1_{c_1}$. Clearly, this resident is either a member of a couple or is a single resident. We now consider both of these cases and show that the we must arrive at the same outcome in either case. In what follows $n_k$ $(1\leq k\leq n_N)$ represents the number of single residents generated following couple $c_k$ as the preference lists of the residents are inferred in the proof.

\noindent Case (i): $r_x$ is single and thus $n_1 > 0$. In this case let $r_x = r^1_{s_1}$. Since $r^1_{s_1}$ is in first place in the preference list of $h^1_{c_1}$, to prevent a fixed assignment, there must exist another hospital that is preferred by $r^1_{s_1}$; let this be $h^2_{c_1}$. If $h^2_{c_1}$ has capacity two then $(r^1_{s_1}, h^2_{c_1})$ represents a fixed assignment, a contradiction. Hence, $h^2_{c_1}$ must have capacity one.

Now, since $r^1_{s_1}$ has $h^2_{c_1}$ in first place in its preference list, there must exist some other resident who is preferred by $h^2_{c_1}$. We consider first the case where each newly generated resident is single. Hence, let this new resident be a single resident, $r^2_{s_1}$. Since $r^2_{s_1}$ is in first place on the preference list of $h^2_{c_1}$ there must exist another hospital which is preferred by $r^2_{s_1}$; let this new hospital be $h^3_{c_1}$. Assume $h^3_{c_1}$ has capacity two. In that case $(r^1_{s_1}, h^2_{c_1})$ represents a fixed assignment, a contradiction. Hence, $h^3_{c_1}$ must have capacity one.

We may continue constructing a sequence of distinct single residents and hospitals of capacity one, but as the number of single residents is finite, ultimately we must eventually arrive at a resident who is a member of a couple; let this resident be $r^1_{c_2}$. Without loss of generality suppose that $r_{c_2}^1$ is the first member of the couple to which he belongs. Let $r^{n_1}_{s_1}$ be the final single resident constructed in the preceding sequence.

It follows that $r^{n_1}_{s_1}$ prefers some hospital $h^{{n_1}+1}_{c_1}$ of capacity one to $h^{n_1}_{c_1}$. If $h^{{n_1}+1}_{c_1} = h^{0}_{c_1}$ then $I$ contains precisely one couple and the instance is of the form shown in Figure \ref{preflists:the full instance} where $N=1$ and $n_1 >0$. Otherwise $h^{{n_1}+1}_{c_1}$ is a new hospital of capacity one that prefers $r^{1}_{c_2}$ to $r^{n_1}_{s_1}$. Since $h^{{n_1}+1}_{c_1}$ has $r^{1}_{c_2}$ in first place on its preference list, it must be the case that $r^{1}_{c_2}$ expresses a joint preference as part of the couple $( r_{c_2}^{1}, r_{c_2}^{2})$ for a hospital pair involving $h^{{n_1}+1}_{c_1}$; let this pair be $( h^{{n_1}+1}_{c_1}, h_{c_2}^{1})$. Since $h^{{n_1}+1}_{c_1}$ has $r^{1}_{c_2}$ in first place on its preference list, $h_{c_2}^{1}$ must be of capacity one and prefer some other resident to $r^2_{c_2}$, otherwise $( r_{c_2}^{1}, r_{c_2}^{2})$ represents a fixed assignment with $( h^{{n_1}+1}_{c_1}, h_{c_2}^{1})$, a contradiction. Now, let this other resident be $r_y$.

\noindent Case (ii): $r_x$ is a member of a couple and thus $n_1 = 0$. Let $r_x = r^1_{c_2}$. Then $h^1_{c_1}$ prefers $r^1_{c_2}$ to $r^2_{c_1}$. Assume that $r^1_{c_2}$ is part of a couple $(r^1_{c_2}, r^2_{c_2})$ and further assume that $(r^1_{c_2}, r^2_{c_2})$ finds $(h^1_{c_1}, h^1_{c_2})$ acceptable. If $h^1_{c_2} = h^0_{c_1}$ then $I$ contains exactly two couples and is of the form shown in Figure \ref{preflists:the full instance} with $N=2$ and $n_1 = n_2 = 0$. (In this case $h^0_{c_1}$ prefers $r^1_{c_1}$ to $r^2_{c_2}$.) Otherwise, $h^1_{c_2}$ is a new hospital which must be of capacity one, or $(r^1_{c_2}, r^2_{c_2})$ represents a fixed assignment with $(h^1_{c_1}, h^1_{c_2})$, and moreover $h^1_{c_2}$ must prefer some other resident to $r^2_{c_2}$; let this resident be $r_y$.

Thus in both cases we have that if $h^1_{c_2} \neq h^0_{c_1}$ then $h_{c_2}^{1}$ is of capacity one and prefers some resident $r_y$ to $r^2_{c_2}$. Clearly, $r_y$ is either a member of a couple or is a single resident. As before, we consider both of these cases and show that we must arrive at the same outcome in either case.

%%%%

\noindent Case (i): $r_y$ is single and thus $n_2 >0$; In this case let $r_y = r^1_{s_2}$. Since $r^1_{s_2}$ is in first place on the preference list of $h^1_{c_2}$, it follows that $h^1_{c_2}$ cannot be in first place in the preference list of $r^1_{s_2}$. Hence, there must exist another hospital preferred by $r^1_{s_2}$; let this be $h^2_{c_2}$. Further, $h^2_{c_2}$ must be of capacity one and have a resident other than $r^1_{s_2}$ in first place in its preference list; let this resident be $r^{2}_{s_2}$. We consider first the case where each newly generated resident is single. Suppose $r^{2}_{s_2}$ is single. Since $r^2_{s_2}$ is in first place on the preference list of $h^2_{c_2}$ there must exist another hospital which is preferred by $r^2_{s_2}$; let this new hospital be $h^3_{c_2}$. Hospital $h^3_{c_2}$ must have capacity one, otherwise $(r^2_{s_2},h^3_{c_2})$ would represent a fixed assignment.

We may continue generating a sequence of distinct single residents and hospitals of capacity one, but since the number of residents is finite, we must eventually arrive at a resident who is a member of a couple; let this resident be $r^1_{c_3}$. Without loss of generality suppose that $r_{c_3}^1$ is the first member of the couple to which he belongs. Let $r^{n_2}_{s_2}$ be the final single resident in the previously generated sequence. Then $r^{n_2}_{s_2}$ prefers some hospital $h^{{n_2}+1}_{c_2}$ to $h^{{n_2}}_{c_2}$ and $h^{{n_2}+1}_{c_2}$ must be of capacity one. If $h^{{n_2}+1}_{c_2} = h^{0}_{c_1}$ then $I$ contains precisely two couples. Otherwise $h^{{n_2}+1}_{c_2}$ is a new hospital of capacity one and prefers $r^1_{c_3}$ to $r^{n_2}_{s_2}$.

Since $h^{{n_2}+1}_{c_2}$ has $r^{1}_{c_3}$ in first place on its preference list, it must be the case that $r^{1}_{c_3}$ expresses a joint preference as part of the couple $( r_{c_3}^{1}, r_{c_3}^{2})$ for a hospital pair involving $h^{{n_2}+1}_{c_2}$; let this pair be $( h^{{n_2}+1}_{c_2}, h_{c_3}^{1})$. 

Since $h^{1}_{c_3}$ has $r^{2}_{c_3}$ in first place on its preference list, $h_{c_3}^{2}$ must be of capacity one and prefer some other resident to $r^2_{c_3}$; let this resident be $r_z$.

\noindent Case (ii): $r_y$ is a member of a couple and thus $n_2 =0$. Let $r_y = r^1_{c_2}$. Then $h^1_{c_2}$ prefers $r^1_{c_3}$ to $r^2_{c_2}$. Assume that $r^1_{c_3}$ is part of a couple $(r^1_{c_3}, r^2_{c_3})$ and further assume that $(r^1_{c_3}, r^2_{c_3})$ finds $(h^1_{c_2}, h^1_{c_3})$ acceptable. If $h^1_{c_3} = h^0_{c_1}$ then $I$ contains three couples and is of the form shown in Figure \ref{preflists:the full instance} with $N=3$ and $n_3 = 0$. (In this case $h^0_{c_1}$ prefers $r^1_{c_1}$ to $r^2_{c_3}$.) Otherwise, $h^1_{c_3}$ is a new hospital which must be of capacity one (or else $(r^1_{c_3}, r^2_{c_3})$ represents a fixed assignment with $(h^1_{c_2}, h^1_{c_3})$ and must prefer some resident to $r^2_{c_3}$; let this resident be $r_z$.

Now, in both cases we have that if $h^1_{c_3} \neq h^0_{c_1}$ then $h^1_{c_3}$ is of capacity one and prefers some resident $r_z$ to $r^2_{c_3}$. As before, we may continue generating a sequence of distinct residents, couples and hospitals in this fashion, but since the number of residents and couples is finite, we must eventually reach some resident, either single or a member of a couple who must be in second place in $h^0_{c_1}$'s preference list and a complete instance of \small$(2,1,2)$\normalsize {\sc -hrc} is formed. Thus, the instance $I$ must be of the form shown in Figure \ref{preflists:the full instance}. \qed
\end{proof}

%Thus, it follows from Lemma \ref{lemma_212HRC_sub-instance} that after all of the fixed assignments have been iteratively satisfied in an instance $I_0$ of \small $(2, 1, 2)$\normalsize {\sc -hrc} the remaining unassigned agents in $I$ will represent the disjoint union of a finite number of discrete sub-instances of \small $(2, 1, 2)$\normalsize {\sc -hrc} of the form shown in Figure \ref{preflists:the full instance}. Let $I' \subseteq I$ be one of these arbitrary disjoint sub-instances of $I$. Whether $I'$ admits a stable matching is determined by the number of couples involved in $I'$. For any $I' \subseteq I$, $I'$ admits a stable matching if and only if the number of couples involved in $I'$ is even. We state this result formally as Lemma \ref{lemma_212HRC_sub-instance_stableifeven}.

%Now, we have that if $I$ is an instance of \small$(2,1,2)$\normalsize {\sc -hrc} in which all fixed assignments have been iteratively satisfied then $I$ is the union of a finite number of disjoint discrete sub-instances of \small$(2,1,2)$\normalsize {\sc -hrc} of the form shown in Figure \ref{preflists:the full instance}. Let $I'$ be one of these arbitrary disjoint sub-instances of $I$. Lemma \ref{lemma_212HRC_sub-instance_stableifeven} proves that $I'$ admits a stable matching if and only if the number of couples involved in $I'$ is even. 

%%%%%%%

\begin{lemma1}
\label{lemma_212HRC_sub-instance_stableifeven}
An instance $I$ of \small$(2,1,2)$\normalsize {\sc -hrc} of the form shown in Figure \ref{preflists:the full instance} admits a stable matching if and only if the number of couples involved in $I$ is even.
\end{lemma1}

\begin{proof}
Let $M$ be a stable matching in $I$. It is either the case that $(r^1_{c_1}, r^2_{c_1})$ is assigned in $M$ or $(r^1_{c_1}, r^2_{c_1})$ is unassigned in $M$. We now consider each of these cases and show that in either case if $I$ contains an odd number of couples then $I$ cannot admit a stable matching.

\noindent Case (i): Assume $(r^1_{c_1}, r^2_{c_1})$ is assigned in $M$ and therefore $(r^1_{c_1}, h^0_{c_1}) \in M$. Clearly either $n_1 = 0$ or $n_1 > 0$. We now show that whether $n_1 = 0$ or $n_1 > 0$, if $(r^1_{c_1}, r^2_{c_1})$ is assigned in $M$ then $(r^1_{c_2}, r^2_{c_2})$ is unassigned in $M$.

%If $n_1 = 0$ and the instance contains exactly one couple then $( r^1_{c_2} , r^2_{c_2})$ is assigned to $( h^1_{c_2} , h^0_{c_1})$ and $h^0_{c_1}$ has $r^2_{c_2}$ in second place on its preference list. However, by assumption $(r^2_{c_1},h^{0}_{c_1}) \in M$, a contradiction. 

If $n_1 = 0$ and the instance contains exactly one couple, then $( r^1_{c_1} , r^2_{c_1})$ represents a fixed assignment with $( h^0_{c_1} , h^1_{c_1})$, a contradiction. Thus, $I$ contains more than one couple. Let the second couple in $I$ be $( r^1_{c_2} , r^2_{c_2})$ such that $h^1_{c_1}$ has $r^1_{c_2}$ in first place on its preference list. We now have that $( r^1_{c_2} , r^2_{c_2})$ expresses a preference for $( h^1_{c_1} , h^1_{c_2})$ and since $(r^{2}_{c_1}, h^{1}_{c_1}) \in M$, clearly $(r^{1}_{c_2}, r^{2}_{c_2})$ cannot be assigned in $M$.

%%%%%

If $n_1 > 0$ then $h^1_{c_1}$ has $r^1_{s_1}$ in first place on its preference list. Now, if $r^1_{s_1}$ is unassigned in $M$ then $(r^1_{s_1}, h^1_{c_1})$ blocks $M$. Hence $r^1_{s_1}$ must be assigned in $M$ and moreover $(r^1_{s_1} , h^2_{c_1}) \in M$. In similar fashion we may confirm that each $r^a_{s_1}$ $(1\leq a < n_1)$ is assigned to the hospital $h^{a+1}_{c_1}$ in first place on its preference list. 

%%%%%

Now consider, $r^{n_1}_{s_1}$. Again $r^{n_1}_{s_1}$ must be assigned to the hospital in first place in its preference list. If $I$ contains exactly one couple then this hospital must be $h^{0}_{c_1}$ by Lemma \ref{lemma_212HRC_sub-instance}. However, by assumption $(r^1_{c_1},h^{0}_{c_1}) \in M$, a contradiction. Thus $I$ must contain more than one couple. Now, let $h^{{n_1}+1}_{c_1}$ be the hospital in first place on $r^{n_1}_{s_1}$'s preference list. Since $(r^{n_1}_{s_1}, h^{{n_1}+1}_{c_1}) \in M$, clearly $(r^{1}_{c_2}, r^{2}_{c_2})$ cannot be assigned in $M$ as the only pair they find acceptable is $(h^{{n_1}+1}_{c_1}, h^1_{c_2})$. Thus, we have that whether $n_1 = 0$ or $n_1 > 0$, if $(r^1_{c_1}, r^2_{c_1})$ is assigned in $M$ then $(r^1_{c_2}, r^2_{c_2})$ is not assigned in $M$.

%%%%%

Now, either $n_2 = 0$ or $n_2 > 0$. We now show that whether $n_2 = 0$ or $n_2 > 0$, if $(r^1_{c_2}, r^2_{c_2})$ is unassigned in $M$ then $(r^1_{c_3}, r^2_{c_3})$ must be assigned in $M$. If $n_2 = 0$ and the instance contains exactly two couples then $( r^1_{c_2} , r^2_{c_2})$ expresses a preference for either $( h^1_{c_1} , h^0_{c_1})$ if $n_1=0$ (or $(h^{{n_1}+1}_{c_1}, h^0_{c_1})$ if $n_1 > 0$) and $h^0_{c_1}$ has $r^2_{c_2}$ in second place on its preference list. In this case, the instance admits exactly two stable matchings of equal cardinality. If $n_2 = 0$ and the instance contains more than two couples then $(r^1_{c_3}, r^2_{c_3})$ expresses a preference for $( h^1_{c_2} , h^1_{c_3})$. Now assume, $h^1_{c_2}$ is unassigned in $M$. Then $(r^1_{c_2}, r^2_{c_2})$ blocks $M$ with $( h^1_{c_1} , h^0_{c_1})$ if $n_1=0$ (or $(h^{{n_1}+1}_{c_1}, h^0_{c_1})$ if $n_1 > 0$), a contradiction. Thus we have that if $(r^1_{c_2}, r^2_{c_2})$ is not assigned in $M$ then $(r^1_{c_3}, r^2_{c_3})$ must be assigned to $( h^1_{c_2} , h^1_{c_3})$ in $M$. 

%%%%%

If $n_2 > 0$ then $h^1_{c_2}$ has $r^1_{s_2}$ in first place on its preference list. Now, if $r^1_{s_2}$ is not assigned in $M$ then $(r^1_{s_2}, h^1_{c_2})$ blocks $M$, a contradiction. Hence $r^1_{s_2}$ must be assigned in $M$ and moreover $(r^1_{s_2} , h^2_{c_2}) \in M$. In similar fashion we may confirm that each $r^{a}_{s_2}$ $(1\leq a \leq n_2)$ must be assigned in $M$ to the hospital $h^{a+1}_{s_2}$ in first place in its preference list.

%%%%% 

Now consider, $r^{n_2}_{s_2}$. If the instance contains exactly two couples then the hospital in first place in the preference list of $r^{n_2}_{s_2}$ must be $h^{0}_{c_1}$ and the result follows. However, if the instance contains more than two couples then the hospital in first place in the preference list of $r^{n_2}_{s_2}$ must be a new hospital $h^{n_2 + 1}_{c_2}$. Now let the next couple be $(r^1_{c_3}, r^2_{c_3})$. Assume $(r^1_{c_3}, r^2_{c_3})$ is unassigned in $M$. Then $(r^{n_2}_{s_2}, h^{n_2 + 1}_{c_2})$ must block $M$, so $(r^1_{c_3}, r^2_{c_3})$ must be assigned to $(h^{n_2 + 1}_{c_2}, h^1_{c_3})$ in $M$.   Thus, whether $n_2 = 0$ or $n_2 > 0$, if $(r^1_{c_2}, r^2_{c_2})$ is unassigned in $M$ then $(r^1_{c_3}, r^2_{c_3})$ must be assigned in $M$.

%%%%% Corrected up to here

%Assume $r^1_{s_3}$ is not assigned in $M$. Thus, $(r^1_{s_3}, h^2_{c_3})$ must block $M$, a contradiction. Thus $r^1_{s_3}$ must be assigned in $M$ and hence $(r^1_{s_3} , h^3_{c_3}) \in M$. In similar fashion we may confirm that each $r^a_{s_3}$ is assigned to the $h^{a+2}_{c_3}$ in first place on its preference list for $1\leq a < n_3$. 

In similar fashion either $n_3 = 0$ or $n_3 >0$. Again, we show that whether $n_3 = 0$ or $n_3 > 0$, if $(r^1_{c_3}, r^2_{c_3})$ is assigned in $M$ then $(r^1_{c_4}, r^2_{c_4})$ is not assigned in $M$. If $n_3 = 0$ and the instance contains exactly three couples then $( r^1_{c_3} , r^2_{c_3})$ is assigned to $( h^1_{c_2} , h^0_{c_1})$ if $n_2 =0$ (or $(h^{{n_2}+1}_{c_1}, h^0_{c_1}$ if $n_2 > 0$) and $h^0_{c_1}$ has $r^2_{c_3}$ in second place on its preference list. However, by assumption $(r^2_{c_1},h^{0}_{c_1}) \in M$, a contradiction. Thus, $I$ contains more than three couples and $(r^1_{c_4}, r^2_{c_4})$ expresses a preference for $( h^1_{c_3} , h^1_{c_4})$ and since $(r^{1}_{c_3}, h^{1}_{c_3}) \in M$, $(r^{1}_{c_4}, r^{2}_{c_4})$ cannot be assigned in $M$.

If $n_3 > 0$ then $h^1_{c_3}$ has $r^1_{s_3}$ in first place on its preference list. Now, if $r^1_{s_3}$ is not assigned in $M$ then $(r^1_{s_3}, h^1_{c_3})$ blocks $M$, a contradiction. Hence $r^1_{s_3}$ must be assigned in $M$ and moreover $(r^1_{s_3} , h^2_{c_3}) \in M$. In similar fashion we may confirm that each $r^a_{s_3}$ $(1\leq a < n_3)$ is assigned to the hospital $h^{a+1}_{c_3}$ in first place on its preference list. 

Now consider $r^{n_3}_{s_3}$. If the instance contains exactly three couples then the hospital in first place in the preference list of $r^{n_3}_{s_3}$ must be $h^{0}_{c_1}$. However, by construction, $(r^2_{c_1}, h^{0}_{c_1}) \in M$, a contradiction. Hence, the instance must have more than three couples and the hospital in first place in the preference list of $r^{n_3}_{s_3}$ must be a new hospital $h^{{n_3}+1}_{c_3}$. Now let the next couple be $(r^1_{c_4}, r^2_{c_4})$.  Since $(r^{n_3}_{s_3}, h^{{n_3}+1}_{c_3}) \in M$, $(r^{1}_{c_4}, r^{2}_{c_4})$ cannot be assigned in $M$.  Thus, whether $n_3 = 0$ or $n_3 > 0$, if $(r^1_{c_3}, r^2_{c_3})$ is assigned in $M$ then $(r^1_{c_4}, r^2_{c_4})$ is not assigned in $M$.
 
%Now assume, $h^2_{c_4}$ is unassigned in $M$. Then $(r^1_{c_4}, r^2_{c_4})$ blocks $M$ with $(h^1_{c_4}, h^2_{c_4})$, a contradiction, thus $h^2_{c_4}$ must be assigned to $r^1_{s_4}$ in $M$. In similar fashion we may confirm that each $r^{a}_{s_4}$ for $1\leq a \leq n_4$ must be assigned in $M$ to the hospital $h^{a+1}_{s_4}$ in second place in its preference list.

Finally we consider whether $n_4 = 0$ or $n_4 > 0$. If $n_4 = 0$ and the instance contains exactly four couples then $( r^1_{c_4} , r^2_{c_4})$ expresses a preference for the hospital pair $( h^1_{c_4} , h^0_{c_1})$ and $h^0_{c_1}$ has $r^2_{c_4}$ in second place on its preference list and the result follows. Otherwise the instance contains more than four couples and $(r^1_{c_5}, r^2_{c_5})$ expresses a preference for $( h^1_{c_4} , h^1_{c_5})$. Now assume, $h^1_{c_4}$ is unassigned in $M$. Then $(r^1_{c_4}, r^2_{c_4})$ blocks $M$ with $(h^{{n_4}+1}_{c_4}, h^1_{c_4})$, a contradiction. Thus $(r^1_{c_5}, r^2_{c_5})$ must be assigned to $( h^1_{c_4} , h^1_{c_5})$ in $M$.

If $n_4 > 0$ then $h^1_{c_4}$ has $r^1_{s_4}$ in first place on its preference list. If $r^1_{s_4}$ is not assigned in $M$ then $(r^1_{s_4}, h^1_{c_4})$ blocks $M$, a contradiction. Hence $r^1_{s_4}$ must be assigned in $M$ and moreover $(r^1_{s_4} , h^2_{c_4}) \in M$. In similar fashion we may confirm that each $r^{a}_{s_4}$ $(1\leq a \leq n_4)$ must be assigned in $M$ to the hospital $h^{a+1}_{s_4}$  in first place in its preference list. Now consider, $r^{n_4}_{s_4}$. If the instance contains exactly four couples then the hospital in first place in the preference list of $r^{n_4}_{s_4}$ must be $h^{0}_{c_1}$ and the result follows. 

At this point we observe that argument is similar for the case that the number of couples is larger than four. As the preceding argument shows, if the number of couples is odd, then no stable matching exists, a contradiction.

\noindent Case (ii): Now suppose that $(r^1_{c_1}, r^2_{c_1})$ is unassigned in $M$. Then essentially $(r^1_{c_1}, r^2_{c_1})$ plays the role of $(r^{1}_{c_2}, r^{2}_{c_2})$ in the proof above and we may continue to generate a sequence of couples, every second of which is unassigned in $M$. Again, the same proof above can be used to infer that if the number of couples is odd, then no stable matching can exist.

\medskip
Conversely, we show that if the number of couples in $I$ is even then $I$ admits a stable matching. For ease of exposition we use the description of the instance $I$ shown in Figure \ref{preflists:the formal instance} for this part of the proof. For clarity, this instance is exactly equivalent to the instance shown in Figure \ref{preflists:the full instance}. Let $M$ be the following matching in $I$ where $h^{{n_N} + 1}_{c_N} = h^0_{c_1}$ if $n_N > 0$ and $h^1_{c_N} = h^0_{c_1}$ if $n_N = 0$:

$
\begin{array}{rcl}
 M & = & \{(r^{1}_{c_1}, h^{0}_{c_1}), (r^{2}_{c_1}, h^{1}_{c_1})\} \\
 & \bigcup & \{(r^1_{c_k}, h^{n_{k-1}+1}_{c_{k-1}}), (r^2_{c_k}, h^1_{c_k}) : 2\leq k\leq N$, $n_{k-1} > 0, k \bmod{2} \neq 0\} \\
 & \bigcup & \{(r^1_{c_k}, h^{2}_{c_{k-1}}), (r^2_{c_k}, h^1_{c_k}) : 2\leq k\leq N$, $n_{k-1} = 0, k \bmod{2} \neq 0\} \\
 & \bigcup & \{(r^{a}_{s_k}, h^{a+1}_{c_k}) : 1\leq k\leq N, 1\leq a\leq n_k , n_k > 0\}	\\
\end{array}
$

Assume $M$ is unstable. Then there must exist a blocking pair of $M$ in $I$. 

Clearly no single resident $r^{a}_{s_k}$ $(1\leq k\leq N, 1\leq a\leq n_k , n_k > 0)$ can form part of a blocking pair for $M$ in $I$ as he is assigned in $M$ to his first preference. Further, no couple $(r^1_{c_k},r^2_{c_k})$ $(2\leq k\leq N, k \bmod{2} \neq 0\}$ can form part of a blocking pair for $M$ in $I$ as they are assigned to the hospital pair in first place on their joint preference list, $(h^{n_{k-1}+1}_{c_{k-1}}, h^1_{c_k})$ if $n_k >0$ or $(h^{2}_{c_{k-1}}, h^1_{c_k})$ if $n_k = 0$.

Now, assume that $(r^1_{c_k}, r^2_{c_k})$ $( 2\leq k\leq N, k \bmod{2} = 1)$ blocks $M$. If $n_{k-1} > 0$ then $(r^1_{c_k}, r^2_{c_k})$ blocks with  $(h^{n_{k-1}+1}_{c_{k-1}}, h^1_{c_k})$. However, $h^{1}_{c_{k}}$ is assigned in $M$ to its first preference $r^{1}_{s_{k}}$ and so cannot form part of a blocking pair, a contradiction. If $n_k =0$ then $(r^1_{c_k}, r^2_{c_k})$ blocks $M$ with $(h^{2}_{c_{k-1}}, h^1_{c_k})$. However, $h^{1}_{c_{k}}$ is assigned in $M$ to its first preference (either $r^{1}_{s_{k}}$ if $n_k >0$, or $r^{1}_{c_{k+1}}$ if $n_k =0$) and so cannot form part of a blocking pair, a contradiction. Since no other possible blocking pairs exist for $M$ in $I$ it must be the case that $M$ is a stable matching in $I$ and the result is proven. \qed
\end{proof}

%Now by Lemma \ref{lemma_212HRC_sub-instance} we have that for any instance $I$ of \small$(2,1,2)$\normalsize {\sc -hrc} in which all fixed assignments have been iteratively satisfied, $I$ is constructed from disjoint sub-instances of the form shown in Figure \ref{preflists:the full instance}. Further, by Lemma \ref{lemma_212HRC_sub-instance_stableifeven} each disjoint sub-instance $I' \subseteq I$ of $I$ admits a stable matching if and only if the number of couples in $I$ is even. However, if every sub-instance $I' \subseteq I$ contains an even number of couples and we assign the residents in $I'$ as in the proof of Theorem \ref{lemma_212HRC_sub-instance_stableifeven} then we form a stable matching in $I$.

%Since all of the stable matchings admitted by $I$ are of the same size, a stable matching is a maximum cardinality stable matching. Hence we may find a maximum cardinality stable matching or report that none exists in an instance \small$(2,1,2)$\normalsize {\sc -hrc} in polynomial time. We state this result formally as Theorem \ref{212HRCEfficientAlgorithmExists}. 
Lemmas \ref{lemma_212HRC_sub-instance} and \ref{lemma_212HRC_sub-instance_stableifeven} lead to the following conclusion.
\begin{theorem1}
\small$(2,1,2)$\normalsize {\sc -min bp hrc} is solvable in polynomial time.
%we can find a maximum cardinality stable matching or report that none exists in polynomial time.
\label{212HRCEfficientAlgorithmExists}
\end{theorem1}
\begin{proof}
Let $I$ be an instance of {\small$(2,1,2)$\normalsize {\sc -hrc}}, and assume that $M_0$ is a matching in $I$ in which all fixed assignments have been iteratively satisfied, and assume that the corresponding deletions have been made from the preference lists in $I$, yielding instance $I'$.  Lemma \ref{lemma_212HRC_sub-instance} shows that $I'$ is a union of sub-instances $I_1,I_2,\dots,I_t$, where each $I_j$ is of the form shown in Figure \ref{preflists:the full instance} ($1\leq j\leq t$).

For each $j$ ($1\leq j\leq t)$, we show how to construct a matching $M_j$ in sub-instance $I_j$ such that $|bp(M_j)|\leq 1$.  Let $N$ be the number of couples in $I_j$.  Suppose firstly that $N$ is even.  The proof of Lemma \ref{lemma_212HRC_sub-instance_stableifeven} shows how to construct a matching $M_j$ that is stable in $I_j$.

Now suppose that $N$ is odd.  Then $N\geq 3$, since all fixed assignments in $I$ have been iteratively satisfied.  By Lemma \ref{lemma_212HRC_sub-instance_stableifeven}, $I_j$ does not admit a stable matching; we will construct a matching $M_j$ in $I$ such that $|bp(M_j)|=1$.  Let $k$ ($1\leq k\leq N$) be given.

Firstly assume that $k$ is odd and $k\neq N$. Match $(r_{c_k}^1,r_{c_k}^2)$ to the hospital pair on their list.  If $n_k>0$, match $r_{s_k}^i$ to his first-choice hospital $h_{c_k}^{i+1}$ ($1\leq i\leq n_k$).  Now assume that $k$ is even.  Leave $(r_{c_k}^1,r_{c_k}^2)$ unassigned.  If $n_k>0$, match $r_{s_k}^i$ to his second-choice hospital $h_{c_k}^i$ ($1\leq i\leq n_k$).  Finally assume that $k=N$.  If $n_N=0$, leave couple $(r_{c_N}^1,r_{c_N}^2)$ unassigned.  Otherwise match $(r_{c_N}^1,r_{c_N}^2)$ to the hospital pair on their list.  Also for each $i$ ($1\leq i\leq n_N-1$), match $r_{s_N}^i$ to his first-choice hospital $h_{c_N}^{i+1}$, and leave $r_{s_N}^{n_N}$ unassigned.

It is straightforward to verify that if $n_N>0$ then $bp(M_j)=\{(r_{s_N}^{n_N},h_{c_N}^{n_N})\}$ in $I_j$.  Otherwise if $n_N=0$ and $n_{N-1}=0$, the only blocking pair of $M_j$ in $I_j$ involves the couple $(r_{c_{N-1}}^1,r_{c_{N-1}}^2)$ and the hospital pair $(h_{c_{N-2}}^1,h_{c_{N-1}}^1)$.  Finally if $n_N=0$ and $n_{N-1}>0$, $bp(M_j)=\{(r_{s_{N-1}}^{n_{N-1}},h_{c_{N-1}}^{n_{N-1}+1})\}$ in $I_j$.

Clearly $M=\cup_{j=0}^t M_j$ is then a most-stable matching in $I$. \qed
\end{proof}

%\begin{corollary1}
%An instance $I$ of \small$(2,1,2)$\normalsize {\sc -hrc} of the form shown in Figure \ref{preflists:the full instance} which does not admit a stable matching can be transformed to an instance which does admit a stable matching by increasing the capacity of any of the hospitals in $I$ by exactly one.
%\label{212-HRC-stableifadding1tohospitalcapacity}
%\end{corollary1}

\section{An Integer Programming formulation for {\sc min bp hrc}}
%\label{section:IPModelsHRC}
\label{section:IPModels_MINBPHRC}
\subsection{Introduction}

%The IP model presented in this section was first published in \cite{McBM13, BMMcB14}.
The IP model for {\sc min bp hrc} is based on modelling the various types of blocking pairs that might arise according to Definition \ref{stability:MM}, and allowing them to be counted by imposing a series of linear inequalities. The variables are defined for each resident, whether single or a member of a couple, and for each element on his preference list (with the possibility of being unassigned). A further consistency constraint ensures that each member of a couple obtains hospitals from the same pair in their list, if assigned.  A suitable objective function then enables the number of blocking pairs to be minimised.  Subject to this, we may also maximise the size of the constructed matching.
%Finally, the objective of the IP is to maximise the size of a stable matching, if one exists.

A crucial component of the IP model is a mapping between the joint preference list of a couple $(r_i,r_j)$ and individual preference lists for $r_i$ and $r_j$; we call these individual lists the \emph{projected preference lists} for $r_i$ and $r_j$.  Let $I$ be an instance of {\sc hrc} with residents $R = \{r_1, r_2,\dots, r_{n_1}\}$ and hospitals $H= \{h_1, h_2,\dots, h_{n_2}\}$. Without loss of generality, suppose residents $r_1, r_2\ldots r_{2c}$ are in couples. Again, without loss of generality, suppose that the couples are $(r_{2i-1}, r_{2i})$  $(1\leq i\leq c)$.  Suppose that the joint preference list of  a couple $\mathcal C_i =  (r_{2i-1}, r_{2i})$ is:
$$\mathcal C_i ~ : (h_{\alpha_1}, h_{\beta_1}),(h_{\alpha_2}, h_{\beta_2})\ldots (h_{\alpha_l}, h_{\beta_l})$$ 
From this list we create the following \emph{projected preference list} for resident $r_{2i-1}$: 
$$r_{2i-1} ~ : ~ h_{\alpha_1}, h_{\alpha_2}\ldots h_{\alpha_l}$$ 
and the following projected preference list for resident $r_{2i}$: 
$$r_{2i} ~ :  ~ h_{\beta_1}, h_{\beta_2}\ldots h_{\beta_l}$$
%Clearly, the projected preference list of the residents $r_{2i-1}$ and $r_{2i}$ are the same length as the preference list of the couple $\mathcal C_i = (r_{2i-1}, r_{2i})$.
Let $l(\mathcal C_i)$ denote the length of the preference list of $\mathcal C_i$ and let $l(r_{2i-1})$ and $l(r_{2i})$ denote the lengths of the projected preference lists of $r_{2i-1}$ and $r_{2i}$ respectively. Clearly we have that $l(r_{2i-1}) = l(r_{2i}) = l(\mathcal C_i)$. A given hospital $h_j$ may appear more than once in the projected preference list of a resident belonging to a couple $\mathcal C_i = (r_{2i-1}, r_{2i})$.

The single residents are $r_{2c+1}, r_{2c+2}\ldots r_{n_1}$, where each such resident $r_i$, has a preference list of length $l(r_i)$ consisting of individual hospitals $h_j\in H$. Each hospital $h_j\in H$ has a preference list of individual residents $r_i\in R$ of length $l(h_j)$. Further, each hospital $h_j\in H$ has capacity $c_j \geq 1$, the maximum number of residents to which it may be assigned.

%When considering the exact nature of a blocking pair in this model, the stability definition due to McDermid and Manlove \cite{MM10} (given in Definition \ref{stability:MM}) is applied in all cases.
%
We describe the variables, constraints and objective function in the IP model $J$ for the {\sc min bp hrc} instance $I$ in Sections \ref{section:IPModelsMINBPHRCVariables}, \ref{section:IPModelsMINBPHRCConstraints} and \ref{sec:objMINBPHRC} respectively.  The text in bold before the definition of a constraint shows the blocking pair type from Definition \ref{stability:MM} to which the constraint corresponds. 
%Hence, a constraint preceded by `\textbf{Stability 1}' is intended to count blocking pairs resulting from Part 1 of the stability definition given by .
Finally in Section \ref{sec:proofMINBPHRC} we present a proof of correctness for the IP model for {\sc min bp hrc}.

\subsection{Variables in the IP model}
\label{section:IPModelsMINBPHRCVariables}
In $J$, for each $i$ $(1\leq i\leq n_1)$ and $p$ $(1\leq p\leq l(r_i))$, define a variable $x_{i,p}$ such that

\[ x_{i,p} = \left\{\begin{array}{ll}
         1 & \mbox{if $r_i$ is assigned to his $p^{th}$ choice hospital}\\
        0 & \mbox{otherwise.}\end{array} \right. \] 

For $p=l(r_i)+1$ define a variable $x_{i,p}$ whose intuitive meaning is that resident $r_i$ is unassigned. Thus we also have that

\[ x_{i,l(r_i)+1} = \left\{\begin{array}{ll}
         1 & \mbox{if $r_i$ is unassigned}\\
        0 & \mbox{otherwise.}\end{array} \right. \] 

Let $X = \{x_{i,p} : 1\leq i \leq n_1 , 1\leq p \leq l(r_i) + 1\}$.  For ease of exposition we define some additional notation.  For given $i$ ($1\leq i\leq n_1)$ and $p$ ($1\leq p\leq l(r_i)$, let $pref(r_i,p)$ denote the hospital at position $p$ of $r_i$'s preference list (this is the projected preference list of $r_i$ if $1\leq i\leq 2c$). %Further let $pref(h_j, q)$ denote the resident at position $q$ of $h_j$'s preference list where $1\leq j\leq n_2$ and $1\leq q\leq l(h_j)$.% 
For each $i$ ($1\leq i\leq c)$ and $p$ ($1\leq p\leq l(r_{2i-1}))$, let $pref((r_{2i-1},r_{2i}),p)$ denote the hospital pair at position $p$ on the joint preference list of $(r_{2i-1},r_{2i})$.

Also, for an acceptable resident-hospital pair $(r_i, h_j)$, let $rank(h_j, r_i) =q$ denote the rank that hospital $h_j$ assigns resident $r_i$, where $1\leq q \leq l(h_j)$. Further, for each $j$ $(1\leq j\leq n_2)$ and $q$ $(1\leq q\leq l(h_j))$ let the set $R(h_j, q)$ contain the resident-position pairs $(r_i, p)$ such that $r_i$ is assigned a rank of $q$ $(1\leq q \leq l(h_j))$ by $h_j$ and $h_j$ is at position $p$ $(1\leq p \leq l(r_i))$ in $r_i$'s preference list (or $r_i$'s projected preference list if $r_i$ belongs to a couple). Hence: 
$$R(h_j, q) = \{(r_i, p)\in R \times \mathbb{Z} :  rank(h_j, r_i) = q \wedge 1\leq p\leq l(r_i)\wedge pref(r_i, p) = h_j\}.$$

Now, for all $j$ $(1\leq j\leq n_2)$ and $q$ $(1\leq q\leq l(h_j))$, define a new variable $\alpha_{j,q} \in \{0,1\}$. The intuitive meaning of a variable $\alpha_{j,q}$ is that if $h_{j}$ is fully subscribed with assignees better than rank  $q$ then $\alpha_{j,q}$ may take the value 0 or 1. However, if $h_{j}$ is not full with assignees better than rank $q$ then $\alpha_{j,q} = 1$. Constraints \eqref{constraint:HRC_MM_ALPHA_DOMAIN} and \ref{definition:alpha} described in Section \ref{section:IPModelsMINBPHRCConstraints} are applied to enforce this property. 

Now, for all $j$ $(1\leq j\leq n_2)$ and $q$ $(1\leq q\leq l(h_j))$, define a new variable $\beta_{j,q} \in \{0,1\}$. The intuitive meaning of a variable $\beta_{j,q}$ is that if $h_{j}$ has $c_j-1$ or more assignees better than rank $q$ then $\beta_{j,q}$ may take a value of 0 or 1. However, if $h_{j}$ has fewer than $c_j-1$ assignees better than rank $q$ then $\beta_{j,q} = 1$. Constraints \eqref{constraint:HRC_MM_BETA_DOMAIN} and \ref{definition:beta} described in Section \ref{section:IPModelsMINBPHRCConstraints} are applied to enforce this property. 

Finally, for all $i$ $(1\leq i\leq n_1)$ and $p$ $(1\leq p\leq l(r_i))$, define a new variable $\theta_{i, p}\in \{0,1\}$.  The intuitive meaning of a variable $\theta_{i,p}$ is that $\theta_{i,p} = 1$ if resident $r_i$ is involved in a blocking pair with the hospital at position $p$ on his preference list, either as a single resident or as part of a couple, and $\theta_{i,p} = 0$ otherwise.
 
\subsection{Constraints in the IP model}
\label{section:IPModelsMINBPHRCConstraints}

The following constraint simply ensures that each variable $x_{i,p}$ must be binary valued for all $i$ $(1\leq i\leq n_1)$ and $p$ $(1\leq p\leq l(r_i)+1)$: 
\begin{equation} \label{constraint:HRC2_1} \displaystyle x_{i,p} \in \{0,1\} \end{equation}
Similarly, the following constraint ensures that each variable $\alpha_{j,q}$ must be binary valued for all $j$ $(1\leq j\leq n_2)$ and $q$ $(1\leq q\leq l(h_j))$:
\begin{equation} \label{constraint:HRC_MM_ALPHA_DOMAIN} \displaystyle \alpha_{j,q} \in \{0,1\} \end{equation}
Also, the following constraint ensures that each variable $\beta_{j,q}$ must be binary valued for all $j$ $(1\leq j\leq n_2)$ and $q$ $(1\leq q\leq l(h_j))$:
\begin{equation}\label{constraint:HRC_MM_BETA_DOMAIN}  \displaystyle \beta_{j,q} \in \{0,1\} \end{equation}
Similarly the following constraint ensures that each variable $\theta_{i,p}$ must be binary valued for all $i$ $(1\leq i\leq n_1)$ and $p$ $(1\leq p\leq l(r_i)+1)$:
\begin{equation}\label{constraint:HRC_MM_THETA_DOMAIN} \displaystyle \theta_{i,p} \in \{0,1\} \end{equation}
As each resident $r_i\in R$ is assigned to exactly one hospital or is unassigned (but not both), we introduce the following constraint for all $i$ $(1\leq i\leq n_1)$: 
\begin{equation} \label{constraint:HRC2_2} \displaystyle \sum\limits_{p=1}^{l(r_i)+1} x_{i,p} = 1 \end{equation}
Since a hospital $h_j$ may be assigned at most $c_j$ residents, $x_{i, p} = 1$ where $pref(r_i,p) = h_j$ for at most $c_j$ residents. We thus obtain the following constraint for all $j$ $(1\leq j\leq n_2)$:
\begin{equation} \label{constraint:HRC2_3} \displaystyle \sum\limits_{i=1}^{n_1} \sum\limits_{p=1}^{l(r_i)} \{x_{i,p} \in X :  pref(r_i, p)=h_j\} \leq c_j \end{equation}
For each couple $(r_{2i-1}, r_{2i})$, $r_{2i-1}$ is unassigned if and only if $r_{2i}$ is unassigned, and $r_{2i-1}$ is assigned to the hospital in position $p$ in their individual list if and only if $r_{2i}$ is assigned to the hospital in position $p$ in their individual list.  We thus obtain the following constraint for all $i$ $(1\leq i\leq c)$ and $p$ $(1\leq p\leq l(r_{2i-1})+1)$:
\begin{equation} \label{constraint:HRC2_4} \displaystyle x_{2i-1,p} = x_{2i,p} \end{equation}
\textbf{Type 1 blocking pairs.} ~ In matching $M$ in $I$, if a single resident $r_i\in R$ is unassigned or has a worse partner than some hospital $h_j\in H$ where $pref(r_i, p)=h_j$ and $rank(h_j, r_i) =q$ then $h_j$ must be fully subscribed with better partners than $r_i$, for otherwise $(r_i,h_j)$ blocks $M$. Hence if $r_i$ is unassigned or has worse partner than $h_j$, i.e., $\sum\limits_{p^{\prime}=p+1}^{l(r_i)+1} x_{i,p^{\prime}}=1$, and $h_j$ is not fully subscribed with better partners than $r_i$, i.e.,  $\sum\limits_{q^{\prime}=1}^{q-1} \{x_{i^{\prime },p^{\prime \prime}} \in X :   ( r_{i^{\prime }}, p^{\prime \prime}) \in R(h_{j}, q^{\prime })\} < c_j$, then we require $\theta_{i,p}=1$ to count this blocking pair. Thus, for each $i$ $(2c+1\leq i\leq n_1)$ and $p$ $(1\leq p\leq l(r_i))$ we obtain the following constraint where $pref(r_i, p) = h_j$ and $rank(h_j, r_i)=q$:
%
%\begin{equation} \label{constraint:HRC2_5} \displaystyle c_j \sum\limits_{p^{\prime}=p+1}^{l(r_i)+1} x_{i,p^{\prime}} \leq \sum\limits_{q^{\prime}=1}^{q-1} \{x_{i^{\prime },p^{\prime \prime}} \in X :   rank(h_j,r_{i^{\prime }}) = q^{\prime } \wedge pref(r_{i^{\prime }}, p^{\prime \prime}) = h_j)\} \end{equation}
%
\begin{equation} \label{constraint:HRC_MINBP_5} \displaystyle  c_j \left(\left(\sum\limits_{p^{\prime}=p+1}^{l(r_i)+1} x_{i,p^{\prime}}\right) - \theta_{i,p}\right)  \leq \sum\limits_{q^{\prime}=1}^{q-1} \{x_{i^{\prime },p^{\prime \prime}} \in X :   ( r_{i^{\prime }}, p^{\prime \prime}) \in R(h_{j}, q^{\prime })\} \end{equation}
In this way, we can count the number of blocking pairs using the $\theta_{i,p}$ values. A similar methodology is used in all replacement constraints for the remaining stability criteria that follow. Ultimately, the number of blocking pairs is the sum of the $\theta_{i,p} $ values, except that to avoid counting a blocking pair twice in the case of a couple, the model will assume that $\theta_{2i,p} = 0$ for all $i$ $(1\leq i\leq c)$ and for all $p$ $(1\leq p\leq l(r_{2i}))$.
\medskip

\noindent
\textbf{Type 2a blocking pairs.} ~ In a matching $M$ in $I$, if a couple $\mathcal C_i=(r_{2i-1}, r_{2i})$ jointly prefer hospital pair $(h_{j_1}, h_{j_2})$, at position $p_1$ in $\mathcal C_i$'s joint preference list, to $( M(r_{2i-1}), M(r_{2i}))$, at position $p_2$, and $h_{j_1}$ is undersubscribed or prefers $r_{2i-1}$ to one of its assignees in $M$, and $h_{j_2} = M(r_{2i})$, then $(r_{2i-1}, r_{2i})$ blocks $M$ with $(h_{j_1}, h_{j_2})$.  In the special case where $pref(r_{2i-1}, p_1)= pref(r_{2i}, p_1) = h_{j_1}$, if $h_{j_1} = h_{j_2} = M(r_{2i})$, $h_{j_1}$ is undersubscribed or prefers $r_{2i-1}$ to one of its assignees in $M$ other than $r_{2i}$, then again $(r_{2i-1}, r_{2i})$ blocks $M$ with $(h_{j_1}, h_{j_2})$.

Thus, for the general case, we obtain the following constraint for all $i$ $(1\leq i\leq c)$ and $p_1, p_2$ $(1\leq p_1 < p_2 \leq l(r_{2i-1}))$ such that $pref(r_{2i}, p_1) = pref(r_{2i}, p_2)$  and $rank(h_{j_1}, r_{2i-1})=q$: 
\begin{equation} \label{constraint:HRC_MINBP_6} \displaystyle  c_{j_1} ( x_{2i-1, p_2} - \theta_{2i-1,p_1}) \leq \sum\limits_{q^{\prime }=1}^{q-1} \{x_{i^{\prime },p^{\prime \prime}} \in X: ( r_{i^{\prime }}, p^{\prime \prime}) \in R(h_{j_1}, q^{\prime })\} \end{equation}
For the special case in which $pref(r_{2i-1}, p_1)= pref(r_{2i}, p_1) = h_{j_1}$ we obtain the following constraint for all $i$ $(1\leq i\leq c)$ and $p_1, p_2$ where $(1\leq p_1 < p_2 \leq l(r_{2i-1}))$ such that $pref(r_{2i}, p_1) = pref(r_{2i}, p_2)$  and $rank(h_{j_1}, r_{2i-1})=q$: 
\begin{multline}
\label{constraint:HRC_MINBP_7} 
\displaystyle ( c_{j_1} - 1) ( x_{2i-1, p_2}  - \theta_{2i-1,p_1}) \leq \\  \sum\limits_{q^{\prime }=1}^{q-1} \{x_{i^{\prime },p^{\prime \prime}} \in X : q^{\prime } \neq rank(h_{j_1}, r_{2i}) \wedge ( r_{i^{\prime }}, p^{\prime \prime}) \in R(h_{j_1}, q^{\prime })\}
\end{multline}
\textbf{Type 2b blocking pairs.} ~ A similar constraint is required for the case that the odd-subscript member of a given couple stays assigned to the same hospital. 
Thus, for the general case, we obtain the following constraint for all $i$ $(1\leq i\leq c)$ and $p_1, p_2$ where $(1\leq p_1 < p_2 \leq l(r_{2i}))$ such that $pref(r_{2i-1}, p_1) = pref(r_{2i-1}, p_2)$  and  $rank(h_{j_2}, r_{2i})=q$:
\begin{equation} \label{constraint:HRC_MINBP_8} \displaystyle  c_{j_2} ( x_{2i-1, p_2}  - \theta_{2i-1,p_1})\leq \sum\limits_{q^{\prime }=1}^{q-1} \{x_{i^{\prime },p^{\prime \prime}}  \in X  : ( r_{i^{\prime }}, p^{\prime \prime}) \in R(h_{j_2}, q^{\prime })\} \end{equation}
Again, for the special case in which $pref(r_{2i-1}, p_1)= pref(r_{2i}, p_1) = h_{j_2}$ we obtain the following constraint for all $i$ $(1\leq i\leq c)$ and $p_1, p_2$ where $(1\leq p_1 < p_2 \leq l(r_{2i}))$ such that $pref(r_{2i-1}, p_1) = pref(r_{2i-1}, p_2)$  and  $rank(h_{j_2}, r_{2i})=q$:
\begin{multline} \label{constraint:HRC_MINBP_9} \displaystyle ( c_{j_1} - 1) ( x_{2i-1, p_2}  - \theta_{2i-1,p_1})\leq \\ \sum\limits_{q^{\prime }=1}^{q-1} \{x_{i^{\prime },p^{\prime \prime}} \in X : q^{\prime } \neq rank(h_{j_2}, r_{2i-1}) \wedge ( r_{i^{\prime }}, p^{\prime \prime}) \in R(h_{j_2}, q^{\prime })\}
\end{multline}
Now, we define a variable $\alpha_{j,q}$ such that if $h_{j}$ is full with assignees better than rank  $q$ then $\alpha_{j,q}$ may take the value of zero or one. Otherwise, $h_{j}$ is not full with assignees better than rank $q$ and $\alpha_{j,q} = 1$. Hence, we obtain the following constraint for all $j$ $(1\leq j\leq n_2)$ and $q$ $(1\leq q\leq l(h_j))$:
\begin{equation} \label{definition:alpha} \displaystyle \alpha_{j,q} \geq  1 - \dfrac { \sum\limits_{q^{\prime }=1}^{q-1} \{x_{i,p} \in X : (r_{i}, p) \in R(h_{j}, q^{\prime }) \} } {c_{j} } \end{equation}
Next we define a variable $\beta_{j,q}$ such that if $h_{j}$ has $c_j-1$ or more assignees better than rank $q$ then $\beta_{j,q}$ may take a value of zero or one. Otherwise, $h_{j}$ has fewer than $c_j-1$ assignees better than rank $q$ and $\beta_{j,q} = 1$. Hence, we obtain the following constraint all $j$ $(1\leq j\leq n_2)$ and $q$ $(1\leq q\leq l(h_j))$:
\begin{equation} \label{definition:beta} \displaystyle \beta_{j,q} \geq 1 - \dfrac { \sum\limits_{q^{\prime }=1}^{q-1} \{x_{i,p} \in X: (r_{i}, p) \in R(h_{j}, q^{\prime }) \}  } { ( c_{j} - 1) } \end{equation}
\textbf{Type 3a blocking pairs.} ~ In a matching $M$ in $I$, if a couple $\mathcal C_i=(r_{2i-1}, r_{2i})$ is unassigned or assigned to a worse hospital pair than $(h_{j_1}, h_{j_2})$ (where $h_{j_1} \neq h_{j_2}$), and for each $t\in \{1,2\}$, $h_{j_t}$ is undersubscribed and finds $r_{2i-2+t}$ acceptable, or prefers $r_{2i-2+t}$ to its worst assignee, then $(r_{2i-1}, r_{2i})$ blocks $M$ with $(h_{j_1}, h_{j_2})$.
%
%
%every post of at least one of $h_{j_1}$ or $h_{j_2}$ has a strictly better partner than $r_{2i-1}$ and $r_{2i}$ respectively. Let $rank(h_{j_1}, r_{2i-1}) = q_1$ and $rank(h_{j_2}, r_{2i}) = q_2$.
%
%
Thus we obtain the following constraint for all $ i$ $(1\leq i\leq c)$ and $p$ $(1\leq p\leq l(r_{2i-1}))$ where $h_{j_1} = pref(r_{2i-1}, p)$, $h_{j_2} = pref(r_{2i}, p)$, $h_{j_1} \neq h_{j_2}$, $rank(h_{j_1}, r_{2i-1}) =q_1$ and $rank(h_{j_2}, r_{2i}) =q_2$:
\begin{equation} \label{constraint:HRC_MINBP_10} \displaystyle \sum\limits_{p^{\prime }=p+1}^{l(r_{2i-1})+1} x_{2i-1, p^{\prime }} + \alpha_{j_1, q_1} + \alpha_{j_2, q_2} - \theta_{2i-1 ,p} \leq 2 \end{equation}
\textbf{Type 3b/c blocking pairs.} ~ In a matching $M$ in $I$, if a couple $\mathcal C_i=(r_{2i-1}, r_{2i})$ is unassigned or assigned to a worse pair than $(h_{j}, h_{j})$ where $M(r_{2i-1})\neq h_j$ and $M(r_{2i})\neq h_j$, $(r_{2i-1}, r_{2i})$ finds $(h_{j}, h_{j})$ acceptable, and $h_{j}$ has two or more free posts available, then $(r_{2i-1}, r_{2i})$ blocks $M$ with $(h_{j}, h_{j})$ -- this is a Type 3b blocking pair. In a matching $M$ in $I$, if a couple $\mathcal C_i=(r_{2i-1}, r_{2i})$ is unassigned or assigned to a worse pair than $(h_{j}, h_{j})$ where $M(r_{2i-1})\neq h_j$ and $M(r_{2i})\neq h_j$, $(r_{2i-1}, r_{2i})$ finds $(h_{j}, h_{j})$ acceptable, and $h_{j}$ prefers at least one of $r_{2i-1}$ or $r_{2i}$ to some assignee of $h_{j}$ in $M$ while simultaneously having a single free post, then $(r_{2i-1}, r_{2i})$ blocks $M$ with $(h_{j}, h_{j})$ -- this is a Type 3c blocking pair.

These two blocking pair types may be modelled by a single constraint. For each $i$ $(1\leq i\leq c)$ and $p$ $(1\leq p\leq l(r_{2i-1}))$  such that $pref(r_{2i-1}, p) = pref(r_{2i},p)$ and $h_j=pref(r_{2i-1}, p)$, where $q = \min \{rank(h_j, r_{2i}),$ $rank (h_j, r_{2i-1})\}$, we enforce the following:
%
%\begin{equation} \label{constraint:HRC2_8} \displaystyle ( c_j-1) \sum\limits_{p^{\prime }=p+1}^{l(r_{2i-1})+1}  x_{2i-1,p^{\prime }} \leq \sum\limits_{q^{\prime }=1}^{l(h_j)} \{x_{i^{\prime },p^{\prime \prime}} : (r_{i^{\prime }}, p^{\prime \prime}) \in S(h_j, q^{\prime })\} \end{equation}
\begin{equation} \label{constraint:HRC_MINBP_11} \displaystyle  c_j \left(\left(\sum\limits_{p^{\prime }=p+1}^{l(r_{2i-1})+1} x_{2i-1,p^{\prime }}\right) - \theta_{2i-1 ,p}\right)  - \dfrac {\sum\limits_{q^{\prime }=1}^{q-1} \{x_{i^{\prime },p^{\prime \prime}} \in X : ( r_{i^{\prime }}, p^{\prime \prime}) \in R(h_{j}, q^{\prime })\}}{ (c_j-1)}  $$ $$ \leq \sum\limits_{q^{\prime }=1}^{l(h_j)} \{x_{i^{\prime },p^{\prime \prime}} \in X: (r_{i^{\prime }}, p^{\prime \prime}) \in R(h_j, q^{\prime })\} \end{equation}
\textbf{Type 3d blocking pairs.} ~ In a matching $M$ in $I$, if a couple $\mathcal C_i=(r_{2i-1}, r_{2i})$ is unassigned or jointly assigned to a worse pair than $(h_{j}, h_{j})$ where $M(r_{2i-1})\neq h_j$ and $M(r_{2i})\neq h_j$, and $h_{j}$ is full and also has two assignees $r_s$ and $r_t$ (where $s\neq t)$ such that $h_{j}$ prefers $r_{2i-1}$ to $r_s$ and $h_j$ prefers $r_{2i}$ to $r_{t}$, then $(r_{2i-1}, r_{2i})$ blocks $M$ with $(h_{j}, h_{j})$.

For each $(h_{j}, h_{j})$ acceptable to $(r_{2i-1}, r_{2i})$, let $r_{min}$ be the better of $r_{2i-1}$ and $r_{2i}$ according to hospital $h_j$ with $rank(h_j, r_{min}) = q_{min}$. Analogously, let $r_{max}$ be the worse of $r_{2i}$ and $r_{2i-1}$ according to hospital $h_j$ with $rank(h_j, r_{max}) = q_{max}$. Then we obtain the following constraint for $i$ $(1\leq i\leq c)$ and $p$ $(1\leq p\leq l(r_{2i-1}))$ such that $pref(r_{2i-1}, p) = pref(r_{2i},p) = h_j$:
\begin{equation} \label{constraint:HRC_MINBP_12} \displaystyle \sum\limits_{p^{\prime }=p+1}^{l(r_{2i-1})+1} x_{2i-1, p^{\prime }} + \alpha_{j, q_{max}} + \beta_{j, q_{min}} - \theta_{2i-1 ,p} \leq 2 \end{equation}
\subsection{Objective function in the IP model}
%\label{section:IPModelsMINBPHRCObjectivefunctions}
\label{sec:objMINBPHRC}
A maximum cardinality most-stable matching $M$ is a matching of maximum cardinality, taken over all most-stable matchings in $I$. To compute such a matching in $J$, we apply two objective functions in sequence.

First we find an optimal solution in $J$ that minimises the number of blocking pairs.  To this end, we apply the following objective function:
%shown in Equation \ref{constraint:HRC_MINBP_13} below.
%A matching $M$ in $I$ with the minimum number of blocking pairs taken over all of the matchings in $I$ requires that the minimum number of $\theta_{i,p}$ must take the value of one. To minimise the sum over all of the values of $i$ and $p$ we apply the following objective function:
\begin{equation} \label{constraint:HRC_MINBP_13} \displaystyle \min \sum\limits_{i=1}^{n_1} \sum\limits_{p=1}^{l(r_i)} \theta_{i,p} \end{equation}

The matching $M$ corresponding to an optimal solution in $J$ will be a most-stable matching in $I$.  Let $k=|bp(M)|$.  Now we week a maximum cardinality matching in $I$ with at most $k$ blocking pairs.  Thus we add the following constraint to $J$, which ensures that, when maximising on cardinality, any solution also has at most $k$ blocking pairs: %returned after finding an optimal solution during the first iteration will be a most-stable matching in $I$. Let $k$ be the number of blocking pairs in $M$. Now we seek a maximum cardinality matching in $I$ with at most $k$ blocking pairs. Thus we apply a constraint that ensures that any solution in the second run also has at most $k$ blocking pairs as follows:
\begin{equation} \label{constraint:HRC_MINBP_14} \displaystyle \sum\limits_{i=1}^{n_1} \sum\limits_{p=1}^{l(r_i)} \theta_{i,p} \leq k \end{equation}

The final step is to maximise the size of the matching, subject to the matching being most-stable.  This involves optimising for a second time, this time using the following objective function:
%A maximum cardinality most-stable matching $M$ is a matching in which the maximum number of residents are assigned in $M$ subject to having the minimum possible number of blocking pairs taken over all of the matchings admitted by $I$. To maximise the size of the matching found, subject to Constraint \eqref{constraint:HRC_MINBP_14} holding, we also apply the following objective function:
\begin{equation} \label{constraint:HRC_MINBP_15} \displaystyle \max \sum\limits_{i=1}^{n_1} \sum\limits_{p=1}^{l(r_i)} x_{i,p}. \end{equation}

\subsection{Proof of correctness for the IP model}
\label{sec:proofMINBPHRC}
We now establish the correctness of the IP model for {\sc min bp hrc} presented in Sections \ref{section:IPModelsMINBPHRCVariables}, \ref{section:IPModelsMINBPHRCConstraints} and \ref{sec:objMINBPHRC}.

\begin{theorem1}\label{MINBPHRC}
Given an instance $I$ of {\sc min bp hrc}, let $J$ be the corresponding IP model as defined in Sections \ref{section:IPModelsMINBPHRCVariables}, \ref{section:IPModelsMINBPHRCConstraints} and \ref{sec:objMINBPHRC} (omitting Constraint \eqref{constraint:HRC_MINBP_14} and objective function \eqref{constraint:HRC_MINBP_15}). A most-stable matching in $I$ is exactly equivalent to an optimal solution to $J$ with respect to objective function \eqref{constraint:HRC_MINBP_13}.
\end{theorem1}
\begin{proof}
Let $M$ be a most-stable matching in $I$. Let $\langle \mbox{\bf x}, \mbox{\boldmath $\alpha$},\mbox{\boldmath $\beta$}\rangle$ be the corresponding assignment of boolean values to the variables in $J$ as constructed in the proof of Theorem 12 in \cite{McBM13}.  Initially let $\theta_{i,p}=0$ for all $i$ ($1\leq i\leq n_1$) and $p$ ($1\leq p\leq l(r_i)$). 

%By Theorem  \ref{IPHRC2}, $r_i$ is involved in a blocking pair with $h_j$, either as a single resident or as part of couple, if and only if a corresponding constraint is violated in the {\sc hrc} model. 
Assume that $(r_i,h_j)$ blocks $M$ where $r_i$ is a single resident and $pref(r_i,p) = h_j$. Then Constraint \eqref{constraint:HRC_MINBP_5} will be violated if $\theta_{i,p} = 0$. Set $\theta_{i,p} = 1$.  Then the LHS of Constraint \eqref{constraint:HRC_MINBP_5} becomes 0 and the constraint is satisfied.

Now, assume that $(r_i,r_j)$ blocks $M$ with $(h_k,h_l)$ for some couple $(r_i,r_j)$, where $pref((r_i,r_j),p) = (h_k,h_l)$.  Then depending on which part of Definition \ref{stability:MM} is violated, one of the constraints in the range \eqref{constraint:HRC_MINBP_6}-\eqref{constraint:HRC_MINBP_9} and
\eqref{constraint:HRC_MINBP_10}-\eqref{constraint:HRC_MINBP_12} will be violated if $\theta_{i,p} = 0$. By setting $\theta_{i,p} = 1$, the constraint concerned will be satisfied.

It follows that $\langle \mbox{\bf x}, \mbox{\boldmath $\alpha$},\mbox{\boldmath $\beta$}, \mbox{\boldmath $\theta$}\rangle$ is a feasible solution to $J$.  Moreover the objective value of this solution is equal to $|bp(M)|$.

\medskip
Conversely, let $\langle \mbox{\bf x}, \mbox{\boldmath $\alpha$},\mbox{\boldmath $\beta$}, \mbox{\boldmath $\theta$}\rangle$ be an optimal solution to $J$ and let $M$ be the corresponding matching in $I$ as constructed in the proof of Theorem 12 in \cite{McBM13}.
%the corresponding assignment of boolean values to the variables in the IP model derived from $I$ as constructed in Section \ref{section:IPModels_MINBPHRC}.

Now assume that $\theta_{i,p} =1$ for some $i$ $(1\leq i\leq n_1)$ and $p$ $(1\leq p\leq l(r_i))$. If $r_i$ is not involved in a blocking pair with $h_j$ where $pref(r_i, p) = h_j$ (either as a single resident or part of a couple), then by Theorem 12 in \cite{McBM13}, Constraints \eqref{constraint:HRC_MINBP_5} to \eqref{constraint:HRC_MINBP_9} and \eqref{constraint:HRC_MINBP_10}-\eqref{constraint:HRC_MINBP_12} are satisfied with $\theta_{i,p} = 0$, in contradiction to the fact that $\langle \mbox{\bf x}, \mbox{\boldmath $\alpha$},\mbox{\boldmath $\beta$}, \mbox{\boldmath $\theta$}\rangle$ is optimal according to objective function \eqref{constraint:HRC_MINBP_13}.  Thus if $\theta_{i,p} = 1$ for some $i$ $(1 \leq i\leq n_1)$ and $p$ $(1\leq p\leq l(r_{n_1}))$ then $r_i$ must be involved in a blocking pair with the hospital in position $p$ on his preference list.

On the other hand, by the first direction, if there is a blocking pair of $M$, there must be a unique corresponding $\theta_{i,p}$ that has value 1.  It follows that $|bp(M)|$ is equal to the objective value of $\langle \mbox{\bf x}, \mbox{\boldmath $\alpha$},\mbox{\boldmath $\beta$}, \mbox{\boldmath $\theta$}\rangle$ in $J$. \qed
%Moreover, if the optimal value of the solution obtained from the model when applying the objective function \ref{constraint:HRC_MINBP_13} for a given model $J$ is $k$, then the minimum number of blocking pairs admitted by any matching in the corresponding {\sc hrc} instance $I$ is $\leq k$. Hence, the result is proven. \qed
\end{proof}

By enforcing Constraint \eqref{constraint:HRC_MINBP_14} and imposing objective function \eqref{constraint:HRC_MINBP_15}, we obtain the following corollary.

\begin{corollary1}\label{MINBPHRC_optimal}
Given an instance $I$ of {\sc min bp hrc} let $J$ be the corresponding IP model as defined in Sections \ref{section:IPModelsMINBPHRCVariables}, \ref{section:IPModelsMINBPHRCConstraints} and \ref{sec:objMINBPHRC} (omitting objective function \eqref{constraint:HRC_MINBP_13}). A maximum cardinality most-stable matching in $I$ is exactly equivalent to an optimal solution to $J$ with respect to objective function \eqref{constraint:HRC_MINBP_15}.
\end{corollary1}

\end{document}